\documentclass{emulateapj} 
\usepackage{apjfonts}
\usepackage{booktabs}
\usepackage{multirow}
\usepackage{xcolor}
\usepackage{soul}
\usepackage{longtable}
 \usepackage{ulem}

\def\cm{cm$^{$-$3}$}

\def\cm3{cm$^{-3}$}

\def\beq{\begin{equation}}
\def\eeq{\end{equation}}
\def\lesssim{\mathrel{\hbox{\rlap{\hbox{\lower4pt\hbox{$\sim$}}}\hbox{$<$}}}}
\def\gtrsim{\mathrel{\hbox{\rlap{\hbox{\lower4pt\hbox{$\sim$}}}\hbox{$>$}}}}

\def\aj{AJ}

\def\apj{ApJ}

\def\apjs{ApJS}
\def\apjl{ApJL}
\def\aap{A\&A}
\def\araa{ARA\&A}

\def\mnras{MNRAS}
\def\nat{Nature}

\def\pasa{Publications of the Astron. Soc. of Australia}

\def\v1d{{\sc v1d}}

\newcommand{\mum}{\ifmmode{\rm \mu m}\else{$\mu$m}\fi}

\slugcomment{\apjl  Accepted, 21 - Feb - 2020}

\shorttitle{CO, water, and methanol in $\eta$~Carinae}
\shortauthors{P. Morris et al.}

\begin{document}

\title{CO, Water, and Possible Methanol in $\eta$ Carinae Approaching Periastron}  

\author{Patrick W. Morris\altaffilmark{1}, 
Steven B. Charnley\altaffilmark{2},
Michael Corcoran\altaffilmark{2,3},
Martin Cordiner\altaffilmark{2,3},\\
Augusto Damineli\altaffilmark{4},
Jose H. Groh\altaffilmark{5}, 
Theodore~R.~Gull\altaffilmark{2}, 
Laurent Loinard\altaffilmark{6},
Thomas Madura\altaffilmark{7},\\
Andrea Mehner\altaffilmark{8},
Anthony Moffat\altaffilmark{9},
Maureen Y. Palmer\altaffilmark{2,3},
Gioia Rau\altaffilmark{2,3},\\
Noel D. Richardson\altaffilmark{10},
and Gerd Weigelt\altaffilmark{11}
 }


\email{pamorris@caltech.edu}

\altaffiltext{1}{California Institute of Technology, IPAC, M/C 100$-$22, Pasadena, CA 91125.}
\altaffiltext{2}{NASA Goddard Space Flight Center, 8800 Greenbelt Road, Greenbelt, MD 20771.}
\altaffiltext{3}{Department of Physics, Catholic University of America, Washington, DC 20064.}
\altaffiltext{4}{Instituto de Astronomia, Geof\'isica e Ci\^encias Atmosf\'ericas da USP, Rua do Mat\~ao 1226, Cidade Universit\'aria S\~ao Paulo-SP, 05508-090, Brasil.}
\altaffiltext{5}{School of Physics, Trinity College Dublin, the University of Dublin, Dublin, Ireland}
\altaffiltext{6}{Instituto de Radioastronomia y Astrofisica, UNAM, Campus Morelia, Michoacan, C.P. 58089, Mexico; Instituto de Astronomia, Universidad Nacional Autonoma de Mexico, Apartado Postal 70-264, 04510 Ciudad de Mexico, Mexico.}
\altaffiltext{7}{Department of Physics and Astronomy, San Jos\'e State University, One Washington Square, San Jose, CA 95192-0106, USA.}
\altaffiltext{8}{European Southern Observatory, Alonso de Cordova 3107, Vitacura, Santiago de Chile, Chile}
\altaffiltext{9}{D\'epartement de Physique, Universit\'e de Montr\'eal, C. P. 6128, succ. centre-ville, Montr\'eal (Qc) H3C 3J7 Centre de Recherche en Astrophysique du Qu\'ebec, Canada}
\altaffiltext{10}{Department of Physics, Embry-Riddle Aeronautical University, 3700 Willow Creek Rd, Prescott, AZ 86301.}
\altaffiltext{11}{Max Planck Institute for Radio Astronomy, Auf dem H\"ugel 69, D-53121 Bonn, Germany.}

\begin{abstract}
 
In circumstellar gas, the complex organic molecule methanol has been found almost exclusively around young stellar objects, and is thus regarded as a signpost of recent star formation.  Here we report the first probable detection of methanol around an evolved high-mass star, in the complex circumstellar environment around the Luminous Blue Variable $\eta$ Carinae, while using ALMA to investigate molecular cloud conditions traced by CO (2-1) in an orbit phase of the massive binary preceding the 2020 periastron.  Favoring methanol over a  $^{13}$CS alternative, the emission originates from hot ($T_{\rm{gas}} \simeq$ 700 K) material, $\sim$2$''$ (0.02 pc) across, centered on the dust-obscured binary in contrast to the CO which traces inner layers of the extended massive equatorial torus, and is accompanied by prominent absorption in a cooler ($T_{\rm{gas}} \simeq$ 110 K) layer of gas.  We also report detections of water in {\it{Herschel}}/HIFI observations at 557 GHz and 988 GHz.  The methanol abundance is several to 50 times higher than observed towards several lower mass stars, while water abundances are similar to those observed in cool, dense molecular clouds.  The very high methanol:water abundance ratio in the core of $\eta$ Carinae may suggest methanol formation processes similar to Fischer-Tropsch-type catalytic reactions on dust grains.  These observations prove that complex molecule formation can occur in the chemically evolved environments around massive stars in the end stages of their evolution, given sufficient gas densities and shielding conditions as may occur in material around massive interacting companions and merger remnants.

\end{abstract}

\keywords{   --- ISM: individual (Homunculus Nebula) --- molecules --- stars: individual ($\eta$~Carinae) }

\section{Introduction}

The formation and survival of dust and molecular gas around hot, massive stars can only take place in regions that can provide sufficient shielding from the intense ionizing stellar radiation fields.  Dust and molecules can be created during the evolution of a single rotating star, or as a consequence of interactions between companions in a multiple star system.   The physical conditions, chemistry, and location of these structures are therefore direct tracers of significant non-spherical mass loss events and stellar wind processes during the evolution of the system.   $\eta$ Car is probably the most luminous star in the Milky Way \citep{1997ARA&A..35....1D,2012MNRAS.423.1623G} and certainly one of the most energetic and best known examples of an evolved system that has undergone multiple high-mass eruptions, traced by its well-recorded light curve variations \citep{2011MNRAS.415.2009S,2019MNRAS.484.1325D}, and the formation of a circumstellar bipolar ``Homunculus'' nebula \citep{2001ApJ...548L.207M} and massive structures in the equatorial plane.  
The equatorial structures, interpreted as a disrupted torus \citep{2018MNRAS.474.4988S}, were detected with mid-infrared imaging \citep{1999Natur.402..502M,1999AJ....118.2369P}, and harbor the bulk of the $\approx$0.25 M$_\odot$ of dust  \citep{2017ApJ...842...79M}, as well as a number of simple H-, C-, N-, and O-bearing molecules with upper level excitation energies to around 300 K \citep{2012ApJ...749L...4L,2016ApJ...833...48L,2017ApJ...842...79M,2018MNRAS.474.4988S,2019MorrInPrep,2019GullInPrep}.   The origin of the torus is not known, but is hypothesized to be the remnant of a merger occurring in the 1840s between massive companions in close binary \citep{2012MNRAS.420.2064M,2015wrs..conf..155M,2017ApJ...842...79M,2016MNRAS.456.3401P,2018MNRAS.480.1466S}.

$\eta$ Car is a present-day binary (i.e., the system may have been born as a triple star system), with a massive ($\simeq$ 120 M$_\odot$) B-type hypergiant undergoing high mass loss from the stellar surface $(\dot{M} = 8.5 \times 10^{-4}$ M$_\odot$ yr$^{-1}$), and an O or Wolf-Rayet type companion of lower mass (25-45 M$_\odot$) and luminosity \citep{2002A&A...383..636P,2010ApJ...710..729M, 2012MNRAS.423.1623G}.  The orbit is highly eccentric ($e$ = 0.9) on a 5.54-year period, separated by 1.54 A.U. at periastron \citep{2008MNRAS.384.1649D,2008MNRAS.386.2330D,2019MNRAS.484.1325D,2009ApJ...707..693M,2016ApJ...819..131T,2017RNAAS...1a...6D} --- about the distance between Mars and the Sun.   Such close proximity between two massive stars results in strong shocks between the colliding winds in a wind-wind collision or WWC region that evolves with orbit phase \citep{2013MNRAS.436.3820M}, and may be responsible for giving the torus a swept out and generally disrupted appearance out to much larger distances.  Surrounding the binary is a knot or shell of dust and gas $< \; 1''$ in diameter (0.01 pc at a distance of 2.35 kpc; Smith 2006), contributing $<$ 10\% to the total emission in the mid-IR \citep{2019A&A...630L...6M}, and Bremsstrahlung and hydrogen recombination line emission \citep{2014ApJ...791...95A}, possibly formed in the WWC.

As part of our studies on the distribution and physical conditions of the shielded regions supporting formation and survival of molecules and dust and thus the mass loss characteristics of $\eta$ Car, we are following the evolution of CO in its lowest rotational energies at different stages of excitation of the nebula, arising as a consequence of the eccentric binary orbit and attenuation of the secondary companion's UV radiation field during periastron passage.  In this paper we present observations of CO $J=2-1$ observed with the Atacama Large Millimeter Array (ALMA) at an orbit phase $\Delta\Phi \simeq$ 0.85, which is at the nebula's highest state of ionization, just prior to periastron passage.  We also present a highly likely detection of methanol at 231.3 GHz, discovered during our investigations with archival ALMA observations to gage variability of CO around periastron passage.  In the standard grain surface chemistry, the formation of methanol should be accompanied by significant levels of water vapor \citep{2002ApJ...571L.173W,2009A&A...505..629F,2015ARA&A..53..541B}.  We report here also detections of water in its two lowest excited states at 557 and 988 GHz in {\it{Herschel}} observations with the Heterodyne Instrument for the Far Infrared (HIFI), where the H$_2$O 1$_{1,0}$-1$_{0,1}$ 557 GHz line is at least a 5-$\sigma$ detection significance. 

While common in the envelopes of young stars, methanol is very rare around more evolved stars, having been reported to date towards only three other systems, all in a post-MS phase:  the binary system HD~101584 in a common envelope phase with an A6Ia primary that terminated as a post-asymptotic giant branch (post-AGB) or red supergiant star (Olofsson et al. 2017, 2019); in the bi-polar nebula around the post-AGB Mira-type variable OH231.8+4.2 \citep{2018A&A...618A.164S}; and a low-significance detection at extremely low abundance towards the massive yellow hypergiant IRC+10420 \citep{2016A&A...592A..51Q}. The detections in $\eta$ Car indicate conditions favorable to complex molecule formation at the upper extreme of stellar mass and evolutionary end point, as a candidate progenitor of a core collapse supernova of Type II (should the primary explode in its current phase) or Type Ib/c.  We report the basic physical parameters indicated by the methanol gas, contrast the location and conditions indicated by $^{12}$CO $J=2-1$ and possibly water, and discuss the implications of a very low water:methanol abundance ratio on the methanol formation pathway.

\begin{figure*}
\begin{center}
\includegraphics[width=16.25cm]{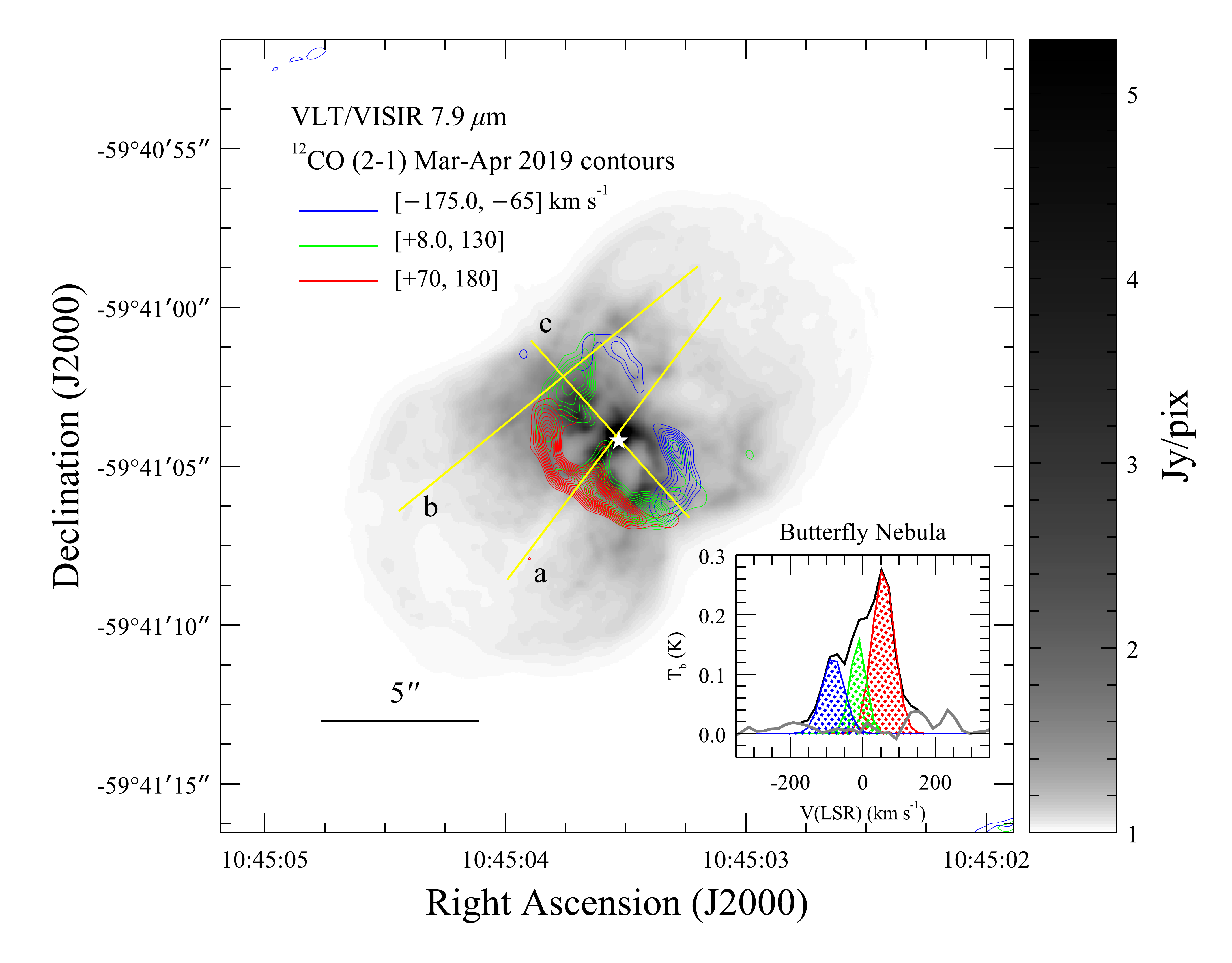} 
\end{center}
\vspace{-1.5em}
 \caption{CO $J = 2-1$ emission in $\eta$ Car.  Contours of CO on a brightness temperature scale (K) are measured over the three principal velocity ranges as indicated, where values in brackets refer to the LSR velocity range of integration.  Contour levels range linearly from 100 to 700 K km s$^{-1}$ in intervals of 50 K km s$^{-1}$. The position of the binary is indicated by the white star.  Yellow lines show the slits used to construct P-V diagrams (Fig~\ref{pvCO}).  The inset shows the average CO emission spectrum with the three principle kinematic components derived by Gaussian deconvolution (contour ranges correspond to each component's FWHM), extracted over a 10$''$ diameter circular aperture centered on the binary's position indicated by the star.  The background image on a reverse log intensity scale is a 7.9 $\mu$m map obtained with VLT/VISIR \cite{2019A&A...630L...6M}.
 \label{CO21}}
\end{figure*} 

{\begin{figure*}
\begin{center}
\includegraphics[width=16.5cm]{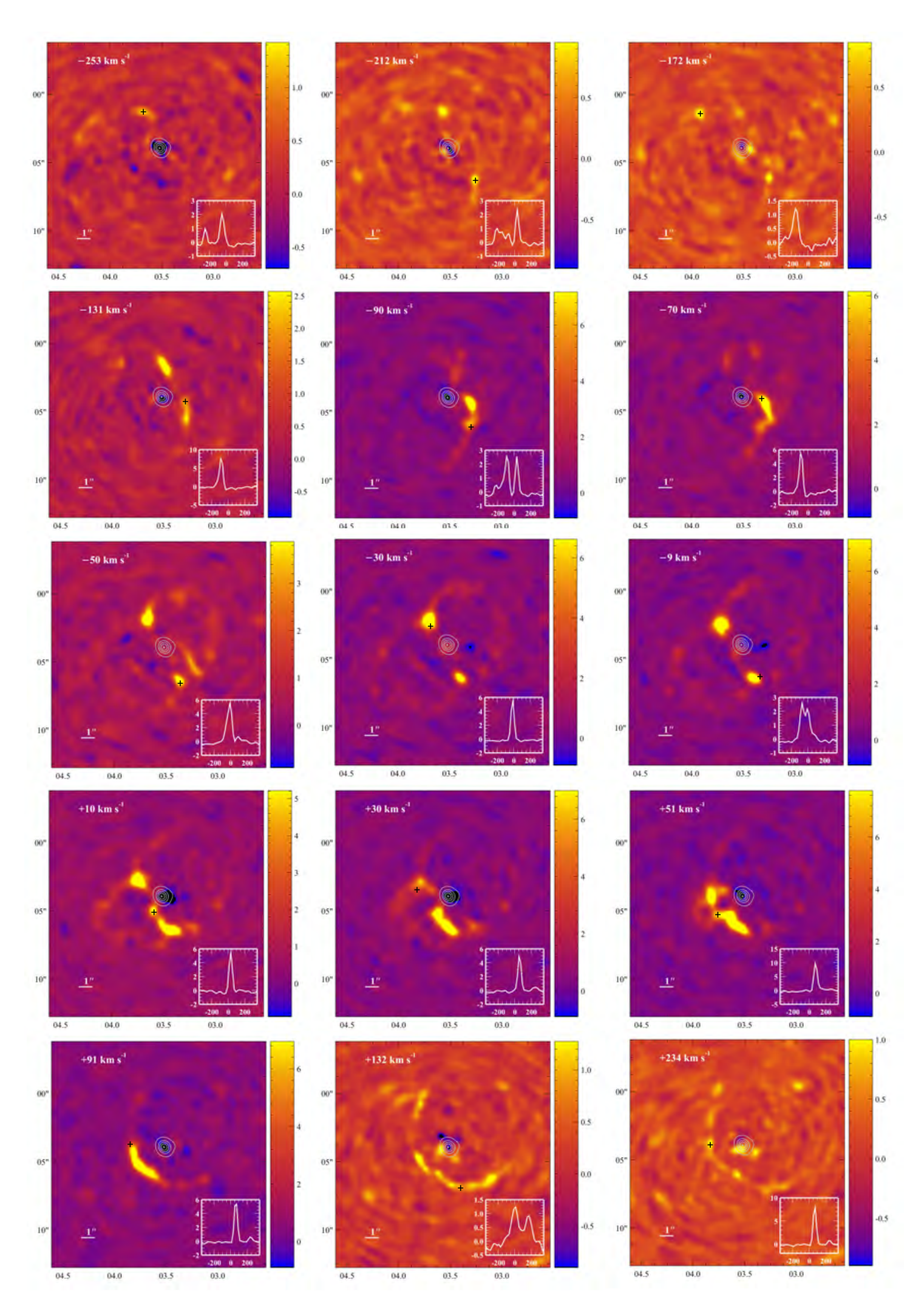}
 \caption{Velocity channels of CO $J=2-1$ observed with ALMA.  The channel maps are plotted on a linear intensity scale in brightness temperature (K).  Coordinates on horizontal and vertical axes are referenced to R.A. 10$^{\rm{h}}$45$^{\rm{m}}$00$^{\rm{s}}$ and decl. $-$59$^\circ$41$'$00$''$, respectively.  Insets show spectral extractions in 2 $\times$ 2 pixels (1.$''$2 $\times$ 1.$''$0) at positions selected for brightness and spectral diversity, indicated by '+'.  The continuum emission at 230 GHz is indicated by the contours on a linear scale in 200 K intervals, centrally peaking at 1200 K.
 \label{COmontage}}
\end{center}
\end{figure*} }

\section{Observations and Analysis Methods}

\subsection{ALMA Observations}

The ALMA data for program 2018.A.00026.S were obtained with Director's Discretionary Time (DDT) in Cycle 6, using 43 to 48 antennae of the 12m main array with a maximum separation of 350m, and the Band 6 receiver tuned to two frequencies, one for each of the $^{12}$CO and $^{13}$CO $J = 2-1$ lines.  A synthesized angular resolution ${\theta}_{\rm{maj}} \times {\theta}_{\rm{min}}$ = 0.$''$58 $\times$ 0.$''$46 was achieved  using two antenna configurations at each tuning, on 2019 March 11 and April 27 for $^{12}$CO, and on 2019 March 29 and April 18 for $^{13}$CO.  The date range corresponds to orbit phase $\Delta\Phi$ = 0.83 - 0.85 in the 5.54-year period, referenced to Epoch 0 defined as periastron on JD 2452819.8 \citep{2008MNRAS.384.1649D}, and was time-constrained in order to measure these lines during the high ionization state of the nebula over a minimal change in orbit phase, prior to the next periastron passage beginning in February 2020.   

The correlators were configured for 1.875 GHz bandwidth ($\approx$2500 km s$^{-1}$) in 128 channels.  The $^{12}$CO tuning included a baseband to observe the H(30)$\alpha$ recombination line.  A continuum baseband was also observed near to each line for continuum subtraction.  Atmospheric variations at each antenna (ranging in 1.1 to 2.4 mm of precipitable water vapor), bandpass, and intensity calibrations were all carried out in CASA 5.4.0-70.   The ALMA calibrator J1032-5917 was used in the bandpass and flux calibrations, yielding an estimated 5\% uncertainty on the intensity scale.  Final image construction in each channel was done using the CASA ``clean'' algorithm on the continuum-subtracted cube.  Brightness temperatures are converted from flux densities $I_\nu$ in mJy/beam using\footnote{https://science.nrao.edu/facilities/vla/proposing/TBconv}

\begin{equation}
T_{\rm{B}} = 1.222 \times 10^3 \frac{I_\nu}{{\nu}^2 {\theta}_{\rm{maj}} {\theta}_{\rm{min}}}, 
\end{equation}

\noindent or $T_{\rm{B}}$ = 4576.1 $I_\nu / {\nu}^2$ in our 2019 DDT observations. 

\subsection{Herschel/HIFI Observations}

The {\it{Herschel}} Space Observatory observations presented here were acquired with the HIFI instrument \citep{1996A&A...315L..49D,2012A&A...537A..17R}) on 02 August 2009 and 31 December 2011, corresponding to phase $\Delta\Phi$ = 0.10 and 0.54 in the preceding oribt cycle.  The 2009 data are from a HIFI observing mode calibration program spectral scan in Mixer Band 1b (obsid 1342181171) over frequencies 555.5 - 593.2 GHz, covering the 556.936 GHz H$_2$O 1$_{1,0}$-1$_{0,1}$ transition, at a frequency redundancy (oversamping rate) of 3.  The standard dual beam switch (DBS) mode with a chopper throw of 3$'$ was employed for baseline corrections.   The 2011 data were obtained in a Band 4a spectral scan from the {\it{Herschel}} Open Time program OT1\_tgull\_3 (T. Gull, P.I.), using the DBS mode (obsid 1342235831) at a frequency redundancy of 4 over the range 950 - 1050 GHz, covering the 987.927 GHz H$_2$O 2$_{0,2}$-1$_{1,1}$ line. The data were processed in the final version 14.1 of the
Standard Product Generation pipeline at the Herschel Science Centre; details of the processing have been given where  the calibration and Open Time observations have been included in a study of $\eta$ Car's mid-IR to millimeter continuum emission  \citep{2017ApJ...842...79M} .   The final data, deconvolved in the pipeline from native dual sideband spectra into single sideband spectra, have a root mean square (RMS) baseline noise of 20 mK in Band 1b and 30 mK in Band 4a, at a smoothed resolution of 2 km s$^{-1}$.  The HIFI half power beam widths ${\theta}_{\rm{beam}}$ are 38.$''$5 at 557 GHz and 21.$''$5 at 998 GHz.  

The low beam filling factors of the central atomic and molecular line emitting region are taken into account  by coupling ${\theta}_s$ to ${\theta}_{\rm{beam}}$ as a function of frequency as follows. The integrated line intensity $I$ of signal expressed on the HIFI antenna temperature $T^*_A$ scales with ${\theta}_s$ as

\begin{equation}
I = \int T_{\rm{B}} dV  =\frac{0.96}{{\eta}_{\rm{beam}}} \int T^*_{\rm{A}} dV \frac{[ {\theta}_{\rm{beam}}^2 + {\theta}_s^2 ]}{ {\theta}_{\rm{beam}}^2 } 
\end{equation}

\noindent where ${\theta}_{\rm{beam}}$ and ${\eta}_{\rm{beam}}$ are the main beam size and efficiency, and the factor 0.96 is the forward efficiency (constant in all HIFI bands).  At our frequencies of interest, ${\eta}_{\rm{beam}}$(557 GHz) = 0.75 and ${\eta}_{\rm{beam}}$(998 GHz) = 0.74.  Cloud source sizes ${\theta}_s$ are constrained by the ALMA CO $J = 2 -1$ and methanol observations presented in this paper, at 0.$''$58 $\times$ 0.$''$46 angular resolution.  A previous spectral analysis of  mid-excitation CO lines observed with HIFI \citep{2017ApJ...842...79M} demonstrated that an emitting source size ${\theta}_s$(CO) $\geq \; 3''$ was required to obtain the best agreement between synthetic and observed line profiles, spanning a range of upper level energies.  In this study we adopt ${\theta}_s$(CO) = 4$''$, better approximating an effective diameter for the location of each of the three principal velocity components of CO $J=2-1$ shown in Figure~\ref{CO21}.  We will later show that ${\theta}_s$(CH$_3$OH) = 1$''$ can be adopted for the methanol gas, although an absorbing region may be slightly more extended than the emitting gas.  For the spatially unresolved HIFI observations we adopt ${\theta}_s$(H$_2$O) = ${\theta}_s$(CH$_3$OH) = 1$''$.  

\subsection{Theoretical methods}

Excitation temperatures $T_{\rm{ex}}$ and column densities $N$ of the molecular emitting regions are calculated using the RADEX radiative transfer code \citep{2007A&A...468..627V}, following an approach similar to that described in our analysis of $^{12}$CO and $^{13}$CO lines observed with HIFI  \citep{2017ApJ...842...79M}.  To summarize, synthetic spectra are computed by solving radiative transfer equations with a number of adjustable input parameters, which are the emitting source size ${\theta}_s$, the molecular excitation temperature $T_{\rm{ex}}$, column density $N$, densities of the collision partners (H$_2$, H, H$^+$, He, and $e^-$), the LSR velocity $v_{\rm{LSR}}$, line width, and continuum levels expressed on a main beam (brightness) temperature scale.  

For abundances relative to H$_2$, a mean column density $N$(H$_2$) = 3.0 $\times$ 10$^{22}$ cm$^{-2}$ is adopted from previous observational work \citep{2006ApJ...644.1151S}.  Unfortunately, observations of H$_2$ at angular scales comparable to our ALMA observations do not yet exist, thus we must regard the relative abundances at a nebular average level, as if all species were observed spatially unresolved.  We might expect some local enhancements of vibrationally-excited H$_2$ in the WWC region and in the dense molecular gas shielded from the WWC, however this could be strongly orbit phase-dependent, and any attempt to estimate variations from the mean would be purely speculative.

The models include treatment for opacity effects in the interaction of each velocity component of each species in the line of sight.  The component with the highest $T_{\rm{ex}}$ is treated as the inner-most, producing a spectrum which is seen as the continuum for the component with the next lowest temperature, and so on.  The code also appropriately broadens the optically thick lines.  $\eta$ Car's far-IR continuum is set at levels measured from the spectral extractions being modeled.  At these wavelengths, the continuum arises primarily from thermal emission by dust grains heated to 95-120 K, and a source of free-free (Bremsstrahlung) radiation at $\lambda \; >$ 300 $\mu$m.  The contributions are readily identified in the full 2.4 $\mu$m to 0.7 mm spectral energy distribution \citep{2017ApJ...842...79M}. Continuum uncertainties including the instrumental baseline RMS noise are taken into account in the measurements of integrated line intensities and fitted output quantities $T_{\rm{ex}}$ and $N$.  At the frequencies of the CO and methanol lines presented in this paper, the molecular gas is exposed more strongly to the free-free photons.  The strength of the free-free emission, originating in a small region of binary stellar wind interactions at the center of the nebula, and the region of molecular line emission may be inter-related through an orbital phase dependency on the [Fe {\sc{ii}}] and [Fe {\sc{iii}}] structures near the binary \citep{2016MNRAS.462.3196G}.  These structures may provide a major source of shielding from the UV radiation field of the companion when it is not submerged in primary star's wind over periastron passage.

We have computed line intensities with combinations of excitation temperature $T_{\rm{ex}}$ and column density $N$ that yield the best agreement with the observed profiles, weighting the fits by the mean RMS noise in the surrounding baseline from each observation.  This is the same as the standard excitation diagram method but does not implicitly rely on an assumption for thermal equilibrium conditions.  For the CO profiles with multiple velocity components, the LSR velocities and line widths are set based on an initial Gaussian deconvolution \citep{2017ApJ...842...79M}, speeding up the overall processing time.  Emission profiles are computed over the temperature range 50 K $\leq$ $T_{\rm{ex}}$ $\leq$ 1200 K in 50 K intervals.  The absorption spectra are fitted over the range of 5 K $\leq$ $T_{\rm{ex}}$ $\leq$ 25 K in 2 K intervals, and  30 K $\leq$ $T_{\rm{ex}}$ $\leq$ 100 K in 5 K intervals.  Initially we assume LTE conditions, and perform cross-checks with the full RADEX code for possible non-LTE effects.  For example, LTE assumes that all level transitions for a given molecule are formed at constant $T_{\rm{ex}}$, and departures could arise at least in part from the effects of a strong continuum, by inducing radiative processes of absorption and stimulated emission in the molecular excitation.   While interactions of the continuum seen by each thermal component of the line emission are included, the radiative coupling is not treated self-consistently with the continuum, in that the RADEX code does not include explicit terms for the continuum in the rate equations for the formation and destruction processes.  

\section{CO (2-1) emission in the Homunculus}

The observed intensity distribution of $^{12}$CO $J=2-1$ emission peaks in three principal velocity ranges, centered at $-$70, $-$5, and $+$70 km s$^{-1}$ ($\pm$10 km s$^{-1}$), spanning a full-width at zero intensity (FWZI) range of $-$150 to $+$150 km s$^{-1}$; see Figure~\ref{CO21}.  At least 7 individual intensity peaks with distinct central velocities and widths 30 $< \; \Delta v \; <$ 70 km s$^{-1}$ have been partially resolved in integrated (single dish) observations of higher-$J$ CO \citep{2012ApJ...749L...4L,2017ApJ...842...79M}.  The kinematic morphology of the molecular gas is presented in Figure~\ref{COmontage} in channels between $-$250 and $+$250 km s$^{-1}$ at a line sensitivity $\sigma_{\rm{rms}}$ of 0.8 mJy over 40 km s$^{-1}$.   A sampling of the emission at locations of maximum brightness temperature $T_{\rm{b}}$ in each channel, converted using 86.1 Jy K$^{-1}$ for the $\approx$0.25 arcsec$^2$ beam, shows a patchy CO line brightness distribution, but the CO velocity field is relatively smooth over the observed range, thus indicating a high degree of coherence in the molecular gas tracing the IR torus.

\begin{figure}
\begin{center}
\includegraphics[width=9.0cm]{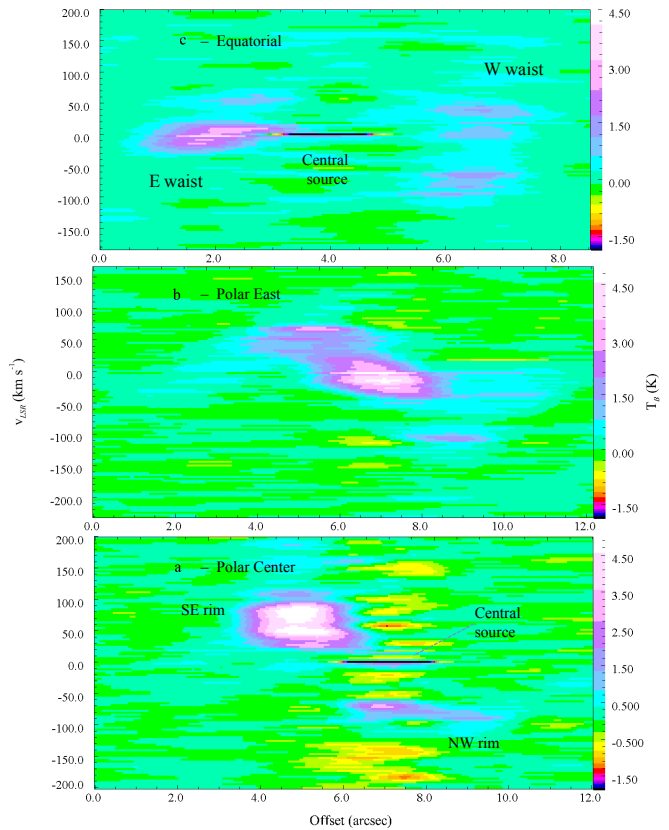}
\end{center}
 \caption{Position-velocity diagrams extracted from slits at the positions indicated in Fig.~\ref{CO21}.  Slits running in the polar direction are 12$''$ long and $\sim$1$''$ wide.  The equatorial slit is 7$'' \times 1''$.  Offsets originate at the labels.  Main features intersected by the slits are indicated.
 \label{pvCO}}
\end{figure} 

The morphology of the CO gas previously observed with ALMA at slightly lower sensitivity (but higher velocity resolution) has been described by \cite{2018MNRAS.474.4988S} as a C-shaped distribution that follows the thermal emission from the disrupted IR torus (referred to as the ``Butterfly Nebula''; \citealt{2005ASPC..332..165C}), with a blow-out region on the blue-shifted near side, NW of the central source where the C opens up (Fig.~\ref{CO21} and \ref{COmontage}).  Position-velocity (P-V) diagrams presented in Figure~\ref{pvCO} generally support this characterization, showing separate red and blue-shifted shells expanding fastest at the rims as measured along the polar axes to the east side of the shell (slit b), and a roughly symmetric, slower velocity field along the equator (slit c).  We note that very little CO emission is detected towards the bright central source, which is marked instead by narrow absorption over the continuum emitting region.  This will be discussed in Sec.~\ref{cold}.   

\begin{figure*}
\begin{center}
\includegraphics[width=10.5cm]{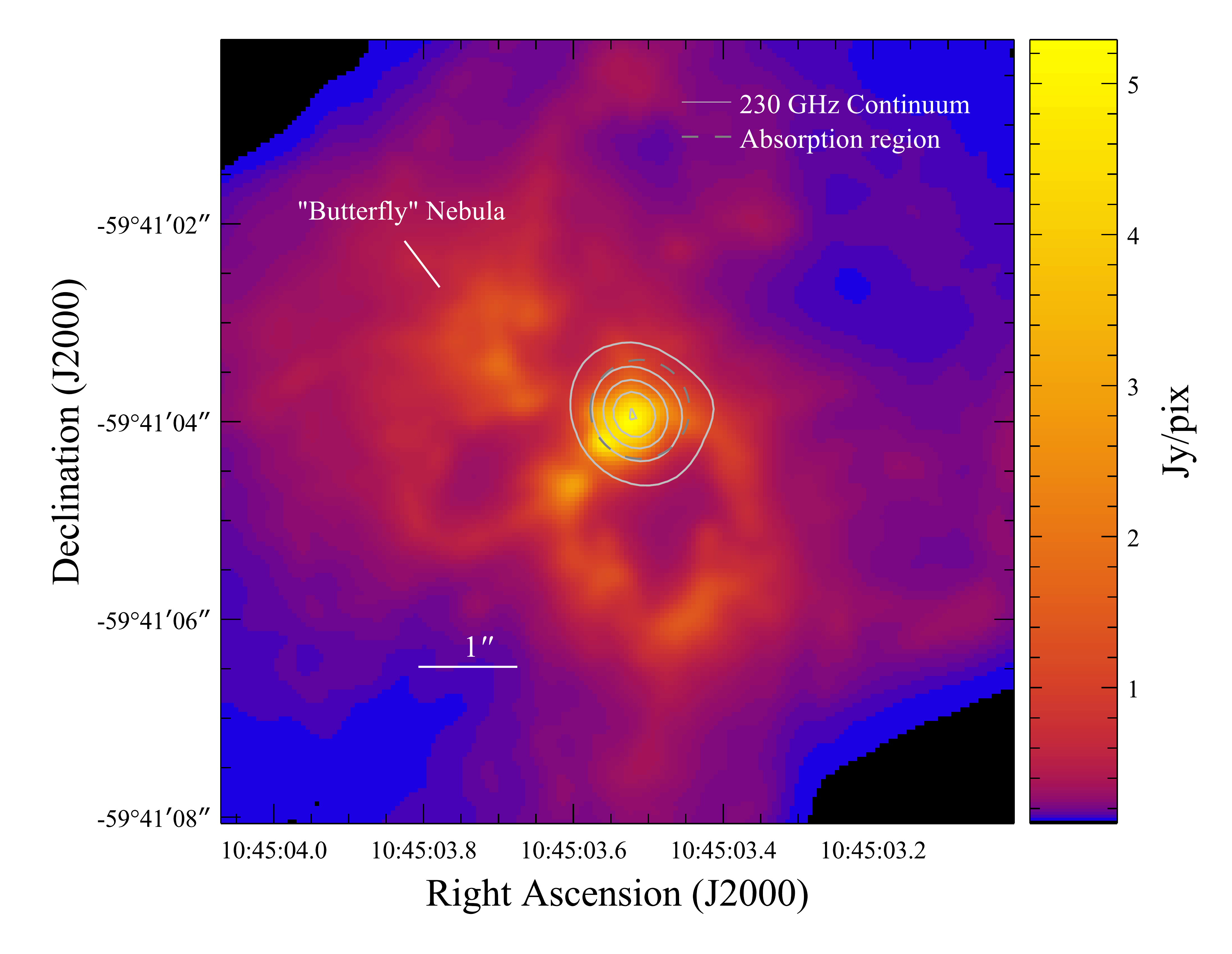}
\includegraphics[width=7.25cm]{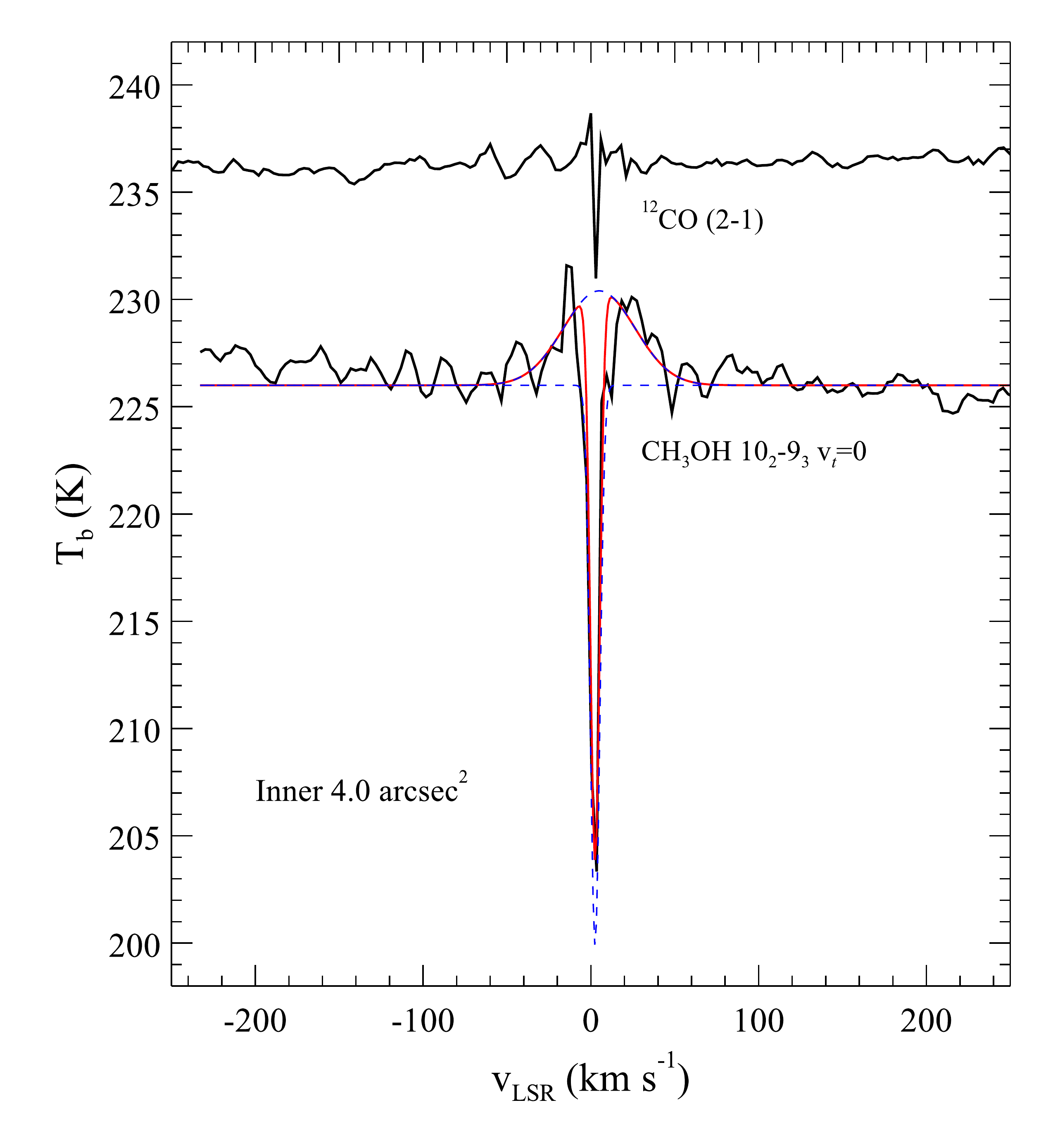} 
\end{center}
 \caption{CO and CH$_3$OH towards the central compact source in $\eta$ Car.  {\em{Left}}:  The inner 8$''$ $\times$ 8$''$ of the 7.9 $\mu$m image shown in Fig.~\ref{CO21}, with contours of 230 GHz continuum emission in brightness temperature overlaid in light gray.  The contours are on a linear scale between 2.0 and 14.0 K in 2.0 K intervals. {\em{Right}}:  Spectral extractions of the CO and methanol features (no continuum subtraction) in circular aperture 1.$'$5 across as indicated in the left plot by the dashed dark gray circle, where methanol is detected to better than 5$\times \sigma_{\rm{RMS}}$ in absorption. Blue dashed lines show synthetic emission and absorption components calculated at the parameter values given in Table~\ref{tableCO}, the red line is the total model.  The CO spectrum has been offset by $+$5 K.
 \label{absPlot}}
\end{figure*} 

The gas is not fully blown out to the NW, however, as described by \cite{2018MNRAS.474.4988S}.  Weak emission to 8 Jy/beam ($T_{\rm{B}}$ = 4.5 K) is observed  in the $-$70 to $-$9 km s$^{-1}$ channels (Fig.~\ref{COmontage} and \ref{pvCO}a), completing two shells of gas each expanding with similar velocity profiles to outer radii of $\sim$2.$''$8.  The remaining molecular gas emission is aligned with the strongest mid-IR emission, tracing thermally protective layers for the enormous amount of dust in the disrupted torus.  Weak knots of emission expanding to 250 km s$^{-1}$ in both directions are detected, coinciding in projection with the interior region of slower moving molecular gas.  No fast clumps or an outward opening angle in the gas are observed in the NW where blow-out has been suggested by \cite{2018MNRAS.474.4988S}.  Nonetheless, the overall CO gas cloud has an orientation that may be related to that of the orbital plane, which is tilted $\sim$45$^\circ$ in the observer's perspective \citep{2012MNRAS.420.2064M,2014MNRAS.442.3316S}.  The orbital angular momentum ($z$) and semi-major ($x$) axes align well with the tenuous NW rim, in the direction where the companion star's wind and radiation field flow over its orbit except when submerged in the primary star's dense wind during periastron passage.  The shape of the CO structure with respect to the binary orbit may be another indication of the companion star's wind influencing the shape of the nebula.

\begin{deluxetable*}{l r r r r r r r r r}
 \tablewidth{0pc}
  \tablecaption{$\eta$ Car CO and methanol gas properties}

\tablehead{
\colhead{Structure\tablenotemark{a}} & & \multicolumn{4}{c}{CO $J = 2-1$} & \multicolumn{4}{c}{CH$_3$OH 10$_2 - 9_3 \; A^-$ $\nu_t$=0} \\
& & \multicolumn{4}{c}{$\nu_0$ = 230.5380 GHz $E_u$ = 16.6 K} & \multicolumn{4}{c}{ $\nu_0$ = 231.2811 GHz $E_u$ =  165.3 K} \\
\cline{3-6} \cline{7-10}  \\
\colhead{} &
\colhead{$V_{\rm{LSR}}$\tablenotemark{b}} &
 \colhead{$\Delta V$} &
\colhead{$I_\nu$\tablenotemark{c}}  &
 \colhead{$T_{\rm{gas}}$} &
 \colhead{$N$(CO)} &
 \colhead{$\Delta V$} &
\colhead{$I_\nu$\tablenotemark{c}}  &
 \colhead{$T_{\rm{gas}}$} &
 \colhead{$N$(CH$_3$OH)} \\
\colhead{} &
 \colhead{(km s$^{-1}$)}  &
 \colhead{(km s$^{-1}$)} & 
 \colhead{(K km s$^{-1}$)} & 
 \colhead{(K)} &
 \colhead{(10$^{16}$ cm$^{-2}$)} &
 \colhead{(km s$^{-1}$)} & 
 \colhead{(K km s$^{-1}$)} & 
 \colhead{(K)} &
 \colhead{(10$^{16}$ cm$^{-2}$)} 
 }

\startdata
Butterfly & $+$72.0 & 55.0 & 17.3 (0.5) & 155 (20) & 4.7 (1.0) & \dotfill & \dotfill & \dotfill & \dotfill \\
 & $+$8.0 & 60.0 & 11.9 (0.5) & 155 (20) & 3.3 (0.8) & \dotfill & \dotfill & \dotfill & \dotfill \\
 & $-$71.0 & 65.0 & 9.3 (0.4) & 155 (20) & 2.9 (0.7) & \dotfill & \dotfill & \dotfill & \dotfill \\
Core & $+$3.0 & 2.50 & $-$17.5 (0.5) & 100 (30) & 1.8 (1.0) & 8.0 & $-$121.0 (3.0) & 110 (30) & 12.0 (4.0)  \\
 & $+$3.0 & \dotfill & \dotfill & \dotfill & \dotfill & 90.0 & 183.0 (5.0) & 700 (100) & 60.0 (15.0) \\
\enddata

\tablenotetext{a}{Spectra for the Butterfly Nebula have been extracted as averages within a circular aperture 10$''$ in diameter, centered on the brightest IR knot coincident with the position of the massive binary. The Core spectra are extracted over an aperture 2$''$ in diameter. }  
\tablenotetext{b}{Local Standard of Rest, $v_{\rm{LSR}} = v - v_0$ where $v_0$ is the line rest frequency measured or calculated in vacuum.  Velocities are not corrected for the system velocity $v_{\rm{sys}} = -8.1 \pm$1 km s$^{-1}$ (heliocentric \citealt{2004MNRAS.351L..15S}).}
\tablenotetext{c}{The line intensity $I_\nu = \int T_{\rm{b}} dv$ integrated over the velocity range of the spectral feature on a brightness temperature scale.  Negative values correspond to line absorption.}
\label{tableCO}

\end{deluxetable*}

The CO $J=2-1$ intensity profile is in good agreement with modeling predictions for $^{12}$CO and $^{13}$CO $J = 5-4$ through $9-8$ lines observed with HIFI \citep{2017ApJ...842...79M}.    The three main velocity components shown in Figure~\ref{CO21} can be reproduced with the same excitation temperatures $T_{\rm{ex}}$ = 135-175 K and column densities $N$(CO) of a few 10$^{16}$. The physical parameters summarized in Table~\ref{tableCO} are based on a representation of the emitting region as a 3$''$-5$''$ diameter sphere.  

An estimate of the mass of the molecular gas around $\eta$ Car based on the CO measurements can be done using the so-called X-factor approach that yields a total molecular hydrogen mass \citep{2013ARA&A..51..207B}; however, the conversion factor $X_{\rm{CO}}$ is considerably uncertain in the case of $\eta$ Car \citep{2018MNRAS.474.4988S}.  Estimates range from 0.2 to 5 M$_\odot$ of H$_2$ at the extremes of the CO/H$_2$ ratios determined for the LBV AG Car, and under typical Milky Way ISM conditions, respectively.   

The total $^{12}$CO $J=2-1$ line intensities integrated over the central 75 arcsec$^2$ in 2015 and 2019 are in close agreement at of 43 $\pm$4 K km s$^{-1}$, taking the different beam sizes of the observations into account in the conversion to $T_B$. We do not necessarily expect emission at the same $J$ level to remain constant over the binary orbit, however, due to expected changes in gas temperatures associated with orbit phase and thus timing of the observations.  The stellar X-ray and UV radiation escaping from the central region steeply declines during periastron passage in reaction to submergence of the hot companion star in the denser primary star's stellar wind \citep{2013MNRAS.436.3820M}, accompanied by a relaxation of ionized systems as radiation is stopped short-ward of 1576 {\AA}  \citep{2012ApJ...746L..18M,2016MNRAS.462.3196G,2017ApJ...838...45C,2018NatAs...2..731H,2019MNRAS.484.1325D}. The minimum in this "light house" effect is reached near orbit phase $\Delta\phi$ = 0.03, and the return to maximum levels occurs by $\Delta\phi \simeq$ 0.15.  The 2019 observations were taken at phase $\Delta\phi$ = 0.85, at maximum excitation of the nebula, while the 2015 spectra were synthesized from observations over 0.08 $\leq \; \Delta\phi \; \leq$ 0.12, as fluences from the escaping UV field are nearly back to their maximum following periastron.  So far we see no measurable response to disturbances in excitation, sampled around the UV light curve minimum. Additional monitoring of the molecular gas during the low ionization state, and over the $\approx$6-7 month field strength recovery phase, are needed to evaluate reaction times and cloud shielding conditions at different locations in the Butterfly, central knot, and larger Homunculus.

\section{Indication of methanol in the inner 0.02 pc}\label{cold}

\subsection{Identification}

We attribute emission and prominent absorption observed at 231.28 GHz to the torsional ground state ($\nu_t$ = 0) transition of CH$_3$OH  10$_2 - 9_3 \; A^-$; see Figure~\ref{absPlot}.  The identification is presently limited to this feature, lacking other significant ($>$ 3-$\sigma_{\rm{RMS}}$) detections in our lower sensitivity HIFI observations.  Nonetheless, it benefits from a lack of line confusion or other suitable atomic or molecular line alternatives within a searched velocity range of $\pm$200 km s$^{-1}$.  Within this range is a rotational transition $^{13}$CS $J=5-4$, discussed below. The remaining transitions are from more complex molecules, most of which have never been observed in space, and most frequently originating from higher upper level energies than those associated with some 15 molecules we have so far identified in $\eta$ Car through multiple transitions \citep{2012ApJ...749L...4L,2019MorrInPrep,2019GullInPrep}.  

The upper level energy of the methanol line $E_u$ = 165 K falls well within the range of other observed  molecular transitions, thus far reaching $E_u$ = 295 K in the N$_2$H$^+$ 11$-$10 observed with HIFI \citep{2019MorrInPrep}. While complex organics would certainly be a surprise in any evolved massive star environment, heavier molecules such as H$^{13}$CN and its tautomer HN$^{13}$C are present \citep{2012ApJ...749L...4L,2019GullInPrep}.  This particular methanol line has been observed in emission in dense molecular clouds, occasionally accompanied by a narrow masing component \citep{2006ARep...50..965S}.   Even at the modest level of excitation of the $J=10-9$ line, the association of methanol with the absorption is also most plausible.  Methanol absorption lines with higher upper level energies have been observed in other environments, such as the $J=5-5$ $K=5-4$ E-type torsional lines (E$_u$ = 171.9 K) observed towards Sgr B2(N) in both emission and significant absorption, tracing warm and very cool (T $\leq$ 10 K) gas clouds \citep{2014ApJ...789....8N}.  We also note the ALMA detection of the ground and first excited states of HCN and H$^{13}$CN $J-4-3$ lines in absorption toward the central star in $\eta$ Car, where the $\nu_2$ = 1 state has an energy of 1050 K, indicating unusually excited conditions close to the binary \citep{2019MNRAS.490.1570B}.

An alternative carrier of the 231 GHz feature may be the $^{13}$CS $J=5-4$ 231.2206 GHz triplet, which would represent the highest molecular mass (45 g/mol) and the first S-bearing molecule detected in $\eta$ Car.  As a first detection of this species, it would also be based on this single feature.  In this case, however, LSR velocities would fall in the range of $-$78 to $-$82 km s$^{-1}$ --- a substantial blue-shift, beyond that of the HCN and H$^{13}$CN absorption lines at $-$60 km s$^{-1}$ reported by \cite{2019MNRAS.490.1570B} and suggested to be associated or entrained with the far-IR continuum-forming region. H recombination emission lines tracing the inner 0.01 pc are blue-shifted by $-$54 to $-$56 km s$^{-1}$ \citep{2014ApJ...791...95A,2020AbrahamInPrep}.  

We performed a search of all HIFI observations of $\eta$ Car, taken in most mixer bands, for transitions of CS as well as OCS, H$_2$CS, and C$_2$S that would support an association of $^{13}$CS with the 231 GHz feature. These are all higher excitation lines, and we limited our search to transitions with upper energies $E_u \; \leq$ 300 K. This produces a list of 94 candidate lines, with $^{12}$CS starting at a rotational excitation of $J=11-10$ and higher.  No features were detected to within 3$\sigma$ of the 10 - 50 mK of baseline noise (varying by band). This does not exclude the presence of CS or other S-bearing molecules, but can be used to set abundance and excitation constraints on required line sensitivities for future searches with ALMA.

The carrier of the 231 GHz feature is unlikely to be located where the H, HCN and H$^{13}$CN are forming in the blue-shifted, kinematically complex Weigelt knots of gas, due to stark differences in the velocity structure and spatial morphology.  The HCN and H$^{13}$CN lines are observed in absorption only towards the central point source, while the emission follows the Butterfly nebula very similar to CO (cf. Fig.~\ref{CO21}) and exhibits a variation in peak intensities with velocity.  Both emission and absorption at 231 GHz, by contrast, follow a significantly wider spatial distribution with little discernible variation in velocity structure, as will be shown in the next section.  The CH$_3$OH identification is a better match to the observations, with the emission well-centered around zero velocity (LSR), consistent with the transitions from 15 or so molecules observed in extended emission and integrated over the inner 50 arcsec$^2$.  We adhere to methanol as the most likely carrier for the remainder of the paper, supported by a tentative (lower significance) detection of methanol emission at HIFI frequencies (included below in Sec.~\ref{watermodel}).  Follow-up observations are needed at other ALMA frequencies to confirm or refute the identification. 

\subsection{Location and characteristics} 

Both emission and absorption components must arise relatively close to the stellar source rather than the foreground ISM, given the upper level energy above the ground state ($E_u$ = 165 K), the line widths (see Table~\ref{tableCO}), and excitation temperatures (discussed below).  Moreover, $^{12}$CO observations of the molecular cloud complex in the Carina H~{\sc{ii}} region by \cite{1998PASA...15..202B} show that there is no detectable interaction between the molecular gas and the Keyhole Nebula (a dense dark cloud NW of $\eta$ Car) or strong radio continuum sources, supporting the idea of a giant molecular cloud (GMC) around the H~{\sc{ii}} region being dispersed by the stellar winds and ionizing fluxes of the massive stars in this region \citep{1995A&A...299..583C}.

Remaining ambiguity regarding the methanol cloud's association with $\eta$ Car can be alleviated by the observed line velocities: the CO emission as well as optically thin H$_2$CO and OH absorption observed at numerous locations in the Carina GMC show the characteristic LSR cloud velocities to be in the range $-$20 to $-$30 km s$^{-1}$ \citep{1973A&A....23...51G,1974A&A....31....5D,1998PASA...15..202B}, whereas the methanol we associate with the 231.3 GHz feature and other molecules observed with $\eta$ Car are symmetrically centered close to zero velocity (LSR). We note in passing that two nearby $E$-symmetry lines at 218.44 and 229.76 GHz were detected serendipitously in the outflow of the low mass Mira-type M star HD~101584 \citep{2017A&A...603L...2O} also while observing $^{12}$CO and $^{13}$CO $J=2-1$ with ALMA.  The 230 GHz line is not detected in  our observations due to poor response at that frequency, near the edge of our baseband set for $^{12}$CO observing.  The 218 GHz line frequency has not yet been covered.

\begin{figure}
\begin{center}
\includegraphics[width=8.75cm]{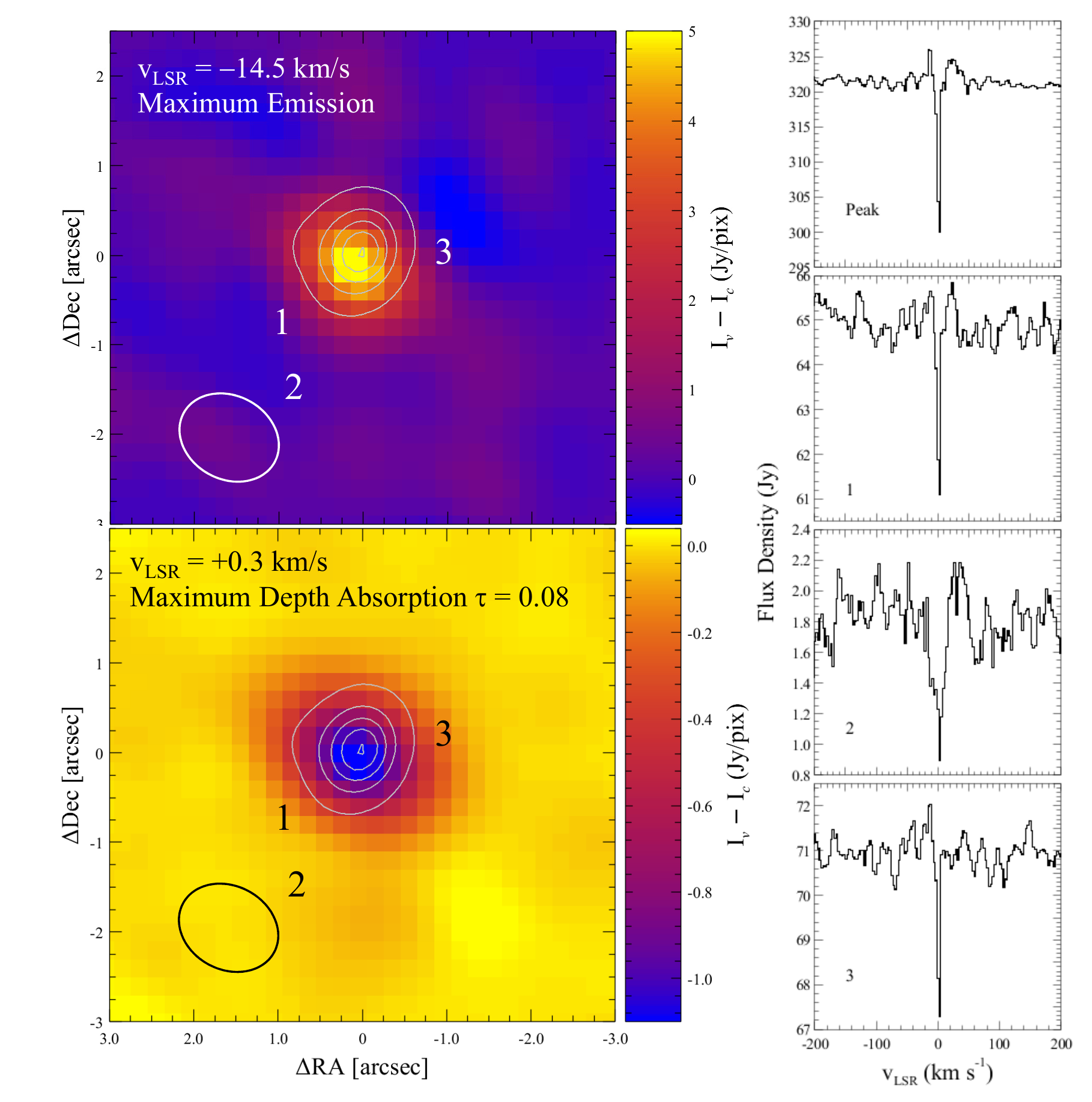}
\end{center}
 \caption{Distribution of 231.3 GHz peak intensity emission and absorption.  Colored maps show velocity channels of maximum emission (left top) and absorption depth (left bottom), using $\nu_0$ = 231.2811 GHz. The continuum in light gray contours is the same as in Fig.~\ref{absPlot} and has been subtracted in both maps.  Numbered positions show where the spectra in the right panel have been sampled over 2 pix $\times$ 2 pix (0.16 arcsec$^2$). The absorption at position 2 is clearly detected but is probably not significant due to the low signal-to-noise ratio. 
 \label{methPlot}}
\end{figure} 

The methanol cloud is compact,  $<$ 2$''$ (0.02 pc) in radius after deconvolving the beam with the intensity distribution of the emission and absorption lines with $>$ 5 $\sigma$ detection, in the velocity channels shown in Figure~\ref{methPlot}.  The emission peak is coincident with the brightest mid-IR knot of emission and peak of the sub-mm continuum, at $T_{\rm{C}}$ = 340 K. Both components are roughly coincident with the 230 GHz continuum, although the absorption can be detected somewhat further out from the central source, up to $\sim$2$''$ at a sensitivity of 0.4 Jy in the 2 pixel $\times$ 2 pixel extractions shown in Figure~\ref{methPlot}.   The localization of the emission and intensity peak on the central source supports the conclusion that the methanol is being excited in proximity to the stellar source.  

At higher angular resolutions, the far-IR/sub-mm continuum has been characterized by Bremsstrahlung radiation from a sub-arcsecond shell radially expanding around the central source at 60 km s$^{-1}$ at the inner part of the shell and 20 km s$^{-1}$ at the outer part \cite{2014ApJ...791...95A}. These characteristics are  consistent with the observed methanol profile, extending over $\pm$(40-50) km s$^{-1}$ in emission, and $\approx$18 km s$^{-1}$ FWZI, or $\approx$8 km s$^{-1}$ full width at half maximum (FWHM) in absorption.  However, as discussed above, variations in the spectral profiles of H recombination emission \citep{2014ApJ...791...95A,2020AbrahamInPrep}, HCN and H$^{13}$CN absorption \citep{2019MNRAS.490.1570B} interpreted to trace the inner continuum-forming region are complex and blue-shifted by tens of km s$^{-1}$, whereas at comparable spectral resolutions, the composite profile of the 231 GHz feature shows practically no discernible variations.   Therefore the relationship of the carrier of the 231 GHz feature to the inner thermal continuum-forming region remains unclear, pending further observations at higher angular resolutions.    

\subsection{Physical conditions}

Estimating the physical conditions of the methanol gas requires assumptions on the total contribution by each of the emission and absorption components to the composite profile shown in Figure~\ref{absPlot}.  We adopt the simplest case of a single emission and single absorption line.  Fitting has been carried out with synthetic spectra computed in non-LTE using the RADEX radiative transfer code \citep{2007A&A...468..627V} with updated collisional rates for CH$_3$OH \citep{2004MNRAS.352...39P}; remaining molecular data are taken from databases at NASA/JPL \citep{1998JQSRT..60..883P}, which have been updated for CH$_3$OH in 2013.  

The best fit to the emission component indicates very warm $T_{\rm{B}}$ = 900 K conditions, with a total column density $N$(CH$_3$OH) $\approx$ 1.5$\times$10$^{-4}$[3.0$\times$10$^{22}$/$N$(H$_2$)] cm$^{-2}$.   This warm component of the gas is a factor 100 more dense than in the nebula of M star HD~101584 \citep{2017A&A...603L...2O}, and nearly four magnitudes higher than towards the yellow hypergiant IRC+10420 \citep{2016A&A...592A..51Q}. If we had adopted two emission components expanding at velocities where the intensities peak at $v_{\rm{LSR}}$ =  $-$15 and $+$25 km s$^{-1}$, the same total column would be summed from red- and blue-shifted gas by a ratio of 1.6:1.  The absorption line has an indicated optical depth $\tau \approx$ 0.08 $\pm$0.01, and can be fit by a cooler $T_{\rm{B}}$ = 110 K layer gas at $\sim$50 times lower density than the warm layer.    

Along the same line of sight to the continuum peak, weak absorption by CO $J=2-1$ to an optical depth $\tau$ $<$ 0.05 is also detected; see Figure~\ref{CO21}.   In contrast with the absorption at 231 GHz, however, the CO absorption line is narrow and unresolved at 2.5 km s$^{-1}$,  probably originating in a foreground ISM cloud.  CO emission tracing the Butterfly Nebula in the disrupted torus is very weak along the peak continuum line of sight.  If located instead in a quiescent layer or shell of gas closer to the compact central source, it could  be related to a compact dusty clump of gas in the line of sight to the binary, gradually dissipating to explain the long-term visual brightening of the central source as proposed by \cite{2019MNRAS.484.1325D}.   Either scenario should be distinguishable by continued monitoring, particular as the stellar binary passes through periastron when variations in excitation of the atomic and molecular gas are expected.  Meanwhile, for a rough measure of the column density assuming $T_{\rm{ex}}$ = 100 K for the absorbing region, $N_{\rm{abs}}$(CO) = 1.8 $\pm$1.0 $\times$10$^{16}$ cm$^{-2}$, a factor 7 lower than $N_{\rm{abs}}$(CH$_3$OH).   

The absorbing layer of CO is not included as an interacting component in our model calculations.  Similar weak absorption of CO $J=4-3$ with a width of 0.6 km s$^{-1}$ along the same line of sight has been reported by \cite{2019MNRAS.490.1570B}, which they also interpret to arise either in a cold foreground cloud or closer to the central source in the dusty clump proposed by \cite{2019MNRAS.484.1325D}.\footnote{Morris et al. (2017) also noted prominent saturated absorption in the CO $J=5-4$ profile observed with HIFI, but attributed this to a self-chopping effect in the dual beam switch observations from extended emission in the Carina Nebula.  The new ALMA observations of CO $J=4-3$ and $3-2$ raise the likelihood that at least some if not most of the CO $J=5-4$ absorption is also occurring in the line of sight.}  Furthermore, we have neglected several potentially important processes in the radiative transfer calculations, such as collisional excitation by He ions \citep{2010MNRAS.403.2033R} and neglect of torsionally-excited states $v_t \; >$ 2  in the rate equations.  The low optical depth of the absorption also increases the sensitivity of the predicted intensities to non-LTE conditions \citep{2016MNRAS.460.2648P}, if the density of hydrogen in this region does not remain significantly (factor of 10 - 100) above the critical density of the transitions observed in absorption.  Nonetheless, these limitations do not alter our main conclusions on the approximate physical conditions and location of the CH$_3$OH carrier of the 231 GHz features, and surrounding CO gas.


\begin{figure}
\begin{center}
\includegraphics[width=8.5cm]{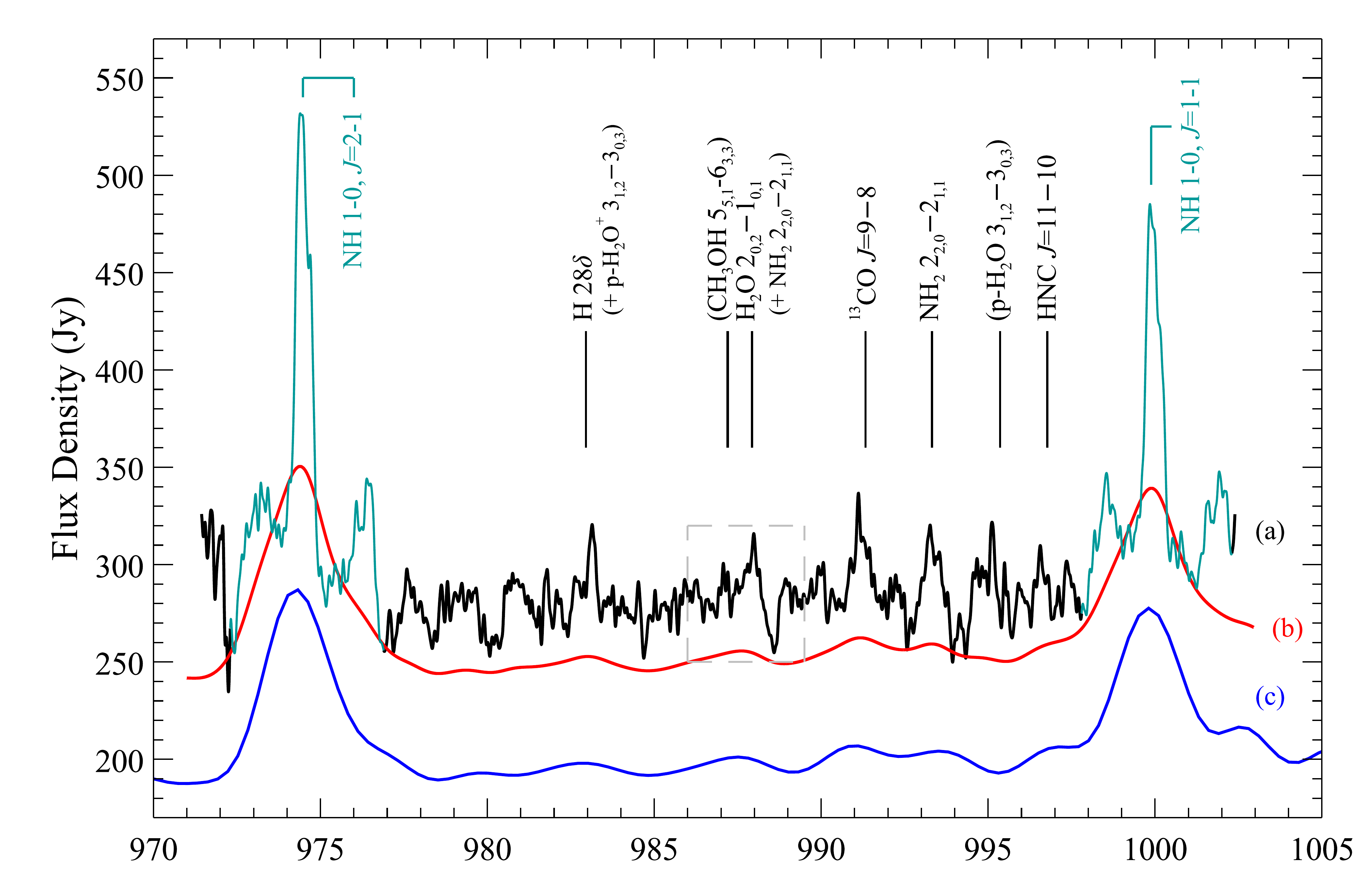}
\includegraphics[width=8.5cm]{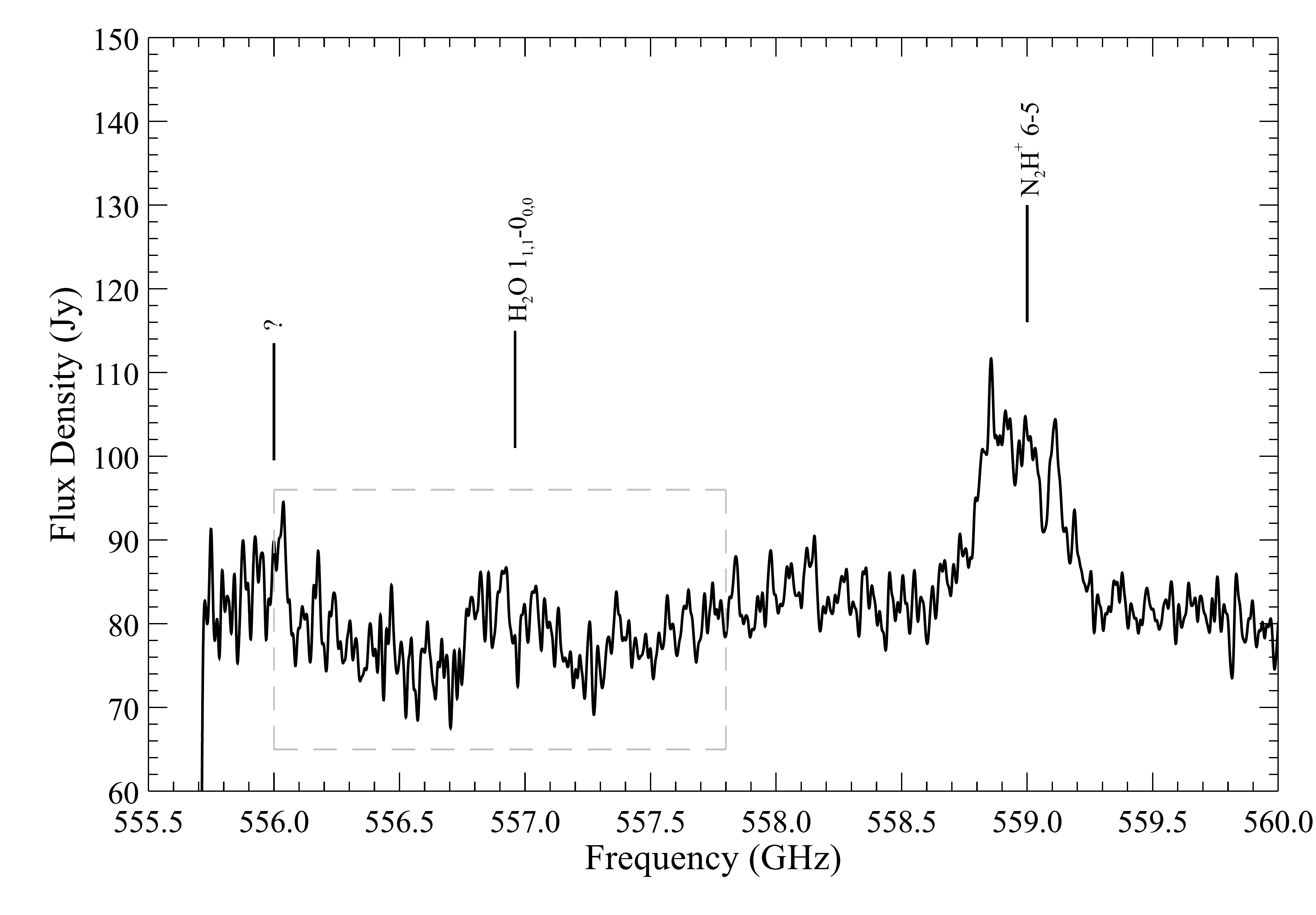}

\end{center}
 \caption{HIFI spectral scans covering the water lines detected at 988 GHz (top) and 557 GHz (bottom).  The top panel shows the HIFI spectrum (a) at a resolution of 2 MHz, and smoothed to 1 GHz resolution (b, in red) comparable to the SPIRE-FTS spectrum (c, in blue).  The smoothed HIFI spectrum has been offset by $-$20 Jy for clarity.  Gray dashed boxes indicate the plotting regions of Fig.~\ref{waterPlot}.  Identifications of the main features are indicated, where those in parentheses are tentative, lacking more than one other transition identified at other frequencies, or very weak and blended. 
 \label{SScansH2O}}
\end{figure}

\section{Water Detection}\label{watersection}

\subsection{Identifications and uncertainties}
 
Figure~\ref{SScansH2O} shows two features in our HIFI spectral scan observations which we identify as H$_2$O 1$_{1,0}$-1$_{0,1}$ 556.936 GHz and 2$_{0,2}$-1$_{1,1}$ 987.927 GHz transitions.  These are {\it{ortho}} (parallel H nuclear spins) and {\it{para}} (opposing) transitions, respectively, and probe cool molecular cloud conditions, arising at 25 K and 55 K above the ground state.  The water emission spans LSR velocities to almost $\pm$150 km s$^{-1}$, about twice as broad as the observed methanol emission at 231 GHz.

Baseline levels are uncertain to 5\% around the 557 GHz line, and 10\% in the spectrum covering the 998 GHz line where line confusion is more apparent.  Identifications of other transitions in both panels are based on connecting lines of the same species observed at other frequencies \citep{2019MorrInPrep,2019GullInPrep}, except those enclosed by brackets which are too close to the baseline uncertainty levels and thus should be regarded as tentative.   Based on narrow ranges of baseline best-guessed to be line-free, the $p$-H$_2$O detection significance is conservatively $\approx$3 $\sigma_{\rm{RMS}}$, but may be slightly higher given the likelihood many other possible weak lines in the spectrum.  This is demonstrated by comparison of our spectrum taken with the Fourier Transform Spectrometer (FTS) on the {\it{Herschel}} Spectral and Photometric Imaging Receiver (SPIRE; \cite{2010A&A...518L...3G}), to the HIFI spectrum smoothed to the comparable resolution of 1.1 GHz.  The correspondence of features is excellent down to the lowest intensity features resolved with SPIRE (to a line:continuum contrast of 2\%), demonstrating that features in the HIFI spectrum at the strength of the 988 GHz line are not instrument artifacts.  The broad, shallow dip around the $o$-H$_2$O (lower panel of Fig.~\ref{SScansH2O}), for which we estimate the detection significance to be 5-6 $\sigma_{\rm{RMS}}$, is likely due baseline drift in the HIFI Band 1b mixer currents over the 556-558 GHz range of LO tunings\footnote{http://herschel.esac.esa.int/twiki/pub/HSC/HiFi/hifi\_handbook.pdf}.  When this happens, the chop rate of the dual beam switched observations taken for baseline calibrations gradually deviates from the Allan stability times at these frequencies.  

Noticable in Figure~\ref{SScansH2O} is a strong hint of absorption resolved near the center of the 557 GHz feature.  At 2 MHz ($\Delta v_{\rm{bin}}$ = 4 km s$^{-1}$) resolution,  $\Delta v_{\rm{FWHM}}$ = 18 km s$^{-1}$ is the full width at half maximum intensity of the spectral line fitted as a Gaussian profile,  and a mean baseline RMS $\sigma_{\rm{RMS}}$ = 0.015 K (see also Fig.~\ref{waterPlot}, lower panel), the absorption line has a significance 3$\sigma$  =  $3 \times \sigma_{\rm{RMS}} \sqrt{2 \Delta v_{\rm{bin}} \Delta v_{\rm{FWHM}}} $ = 0.52 K km s$^{-1}$, compared to the measured value of 0.5 $\pm$0.1 K km s$^{-1}$.

This source of this component cannot be directly localized in the large HIFI beam to assess whether it arises in the compact continuum line of sight or in a more extended self-absorbing layer.  The 231 GHz absorption which we associate with methanol is offset by $+$20 km s$^{-1}$, and is narrower by 10 km s$^{-1}$, while the H$_2$O absorbs at 2-3 times higher optical depth, $\tau$ = 0.25 $\pm$ 0.05.   Within 10 - 1000 AU (< 0.$''$3) of the central star and expanding outward at 40 - 50 km s$^{-1}$ are the so-called ``Weigelt knots'' of dense slow-moving condensations of gas \citep{1986A&A...163L...5W,2004MNRAS.351L..15S} exhibiting numerous ionized atomic emission and absorption lines.  The lines increase in strength towards the UV \citep{2012A&A...540A.133Z} with a peculiar ``inverted'' ionization profile \citep{2013ApJ...773...27R}. HCO$^+$ $J = 4-3$ emission observed with ALMA at velocity $v_{\rm{LSR}}$ $\approx$ $-$12 km s$^{-1}$, similar to the H$_2$O 557 GHz absorption, is found to extend to at least one of these clumps \citep{2019MNRAS.490.1570B}.  The high ionization state of the Weigelt knots makes it unlikely that molecules can survive there, however, thus the H$_2$O absorption is not reliably associated with any specific physical features in the interior based solely on its velocities.

\subsection{Synthetic profiles}\label{watermodel}

\begin{figure}
\begin{center}
\includegraphics[width=8.5cm]{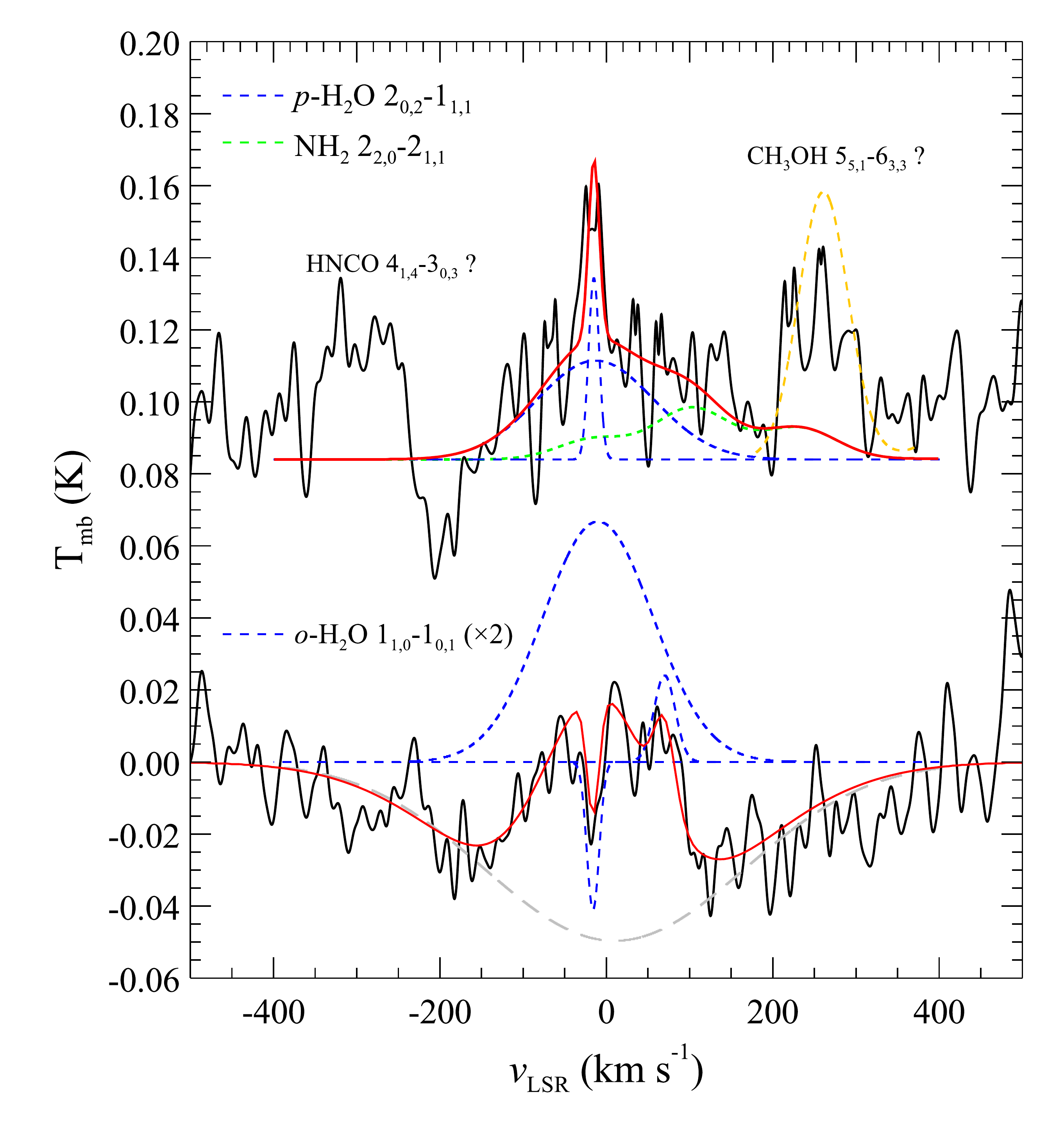}
\end{center}
 \caption{Synthetic profile fits to H$_2$O 1$_{1,0}$-1$_{0,1}$ and 2$_{0,2}$-1$_{1,1}$ in $\eta$ Car.    Dashed lines are predictions for each velocity component in this interacting model (see Table~\ref{water}), red lines show the total fit. The weak NH$_2$ contribution in the red wing of the p-H$_2$O line is a prediction based on modeling of nitrogen hydrides observed with HIFI \citep{2019MorrInPrep}.  The labeled HNCO and CH$_3$OH are tentative identifications.  The orange dashed line is the CH$_3$OH 5$_{5}$-6$_{3}$ $E$ profile predicted from the model for the 10$_2$-9$_3$ $A^-$ line (Fig.~\ref{absPlot} and Table~\ref{tableCO}). For clarity the upper spectrum has been offset by +85 mK, and the lower spectrum has been scaled by a factor 2.  The continuum has been subtracted, and data have been smoothed to 4 km s$^{-1}$.  
 \label{waterPlot}}
\end{figure}

The $o$- and $p-$H$_2$O transitions each have their own energy scheme, thus the cloud conditions they trace should be analyzed independently.  For lack of additional lines to constrain the excitation temperatures, we first compute synthetic profiles of the 557 GHz line using the RADEX code over a range of $N$($o$-H$_2$O) and $T_{\rm{ex}}$, and then we adopt the same $T_{\rm{ex}}$ range on the assumption that both lines are formed in the same volume to estimate $N$($p$-H$_2$O) accordingly.  The predictions at the best-fit temperatures and densities listed in Table~\ref{water} are shown in Figure~\ref{waterPlot}.  The indicated abundances of $\sim$1 $\times$ 10$^{-9}$ [3.0$\times$10$^{22}$/$N$(H$_2$)] cm$^{-2}$ in $\eta$ Car are very low, much lower than in warm star forming envelopes, but in the range of cool ($<$ 100 K), dense molecular clouds where freeze-out takes place \citep{2003A&A...406..937B,2000ApJ...539L.101S,2002ApJ...581L.105B,2012A&A...542A..76H,2013ApJ...765...61E}.  At such low abundances a photon-induced water chemistry is favored over formation in hot, shocked gas.

The  water model includes a narrow component at $v_{\rm{LSR}}$ = $-$15/$-$16 km s$^{-1}$, $\Delta v$ $\approx$ 15-20 km s$^{-1}$, fitting the  absorption at 557 GHz using the parameters listed in Table~\ref{water}.  This model predicts the $p$-H$_2$O $2_{0,2} - 1_{1,1}$ in emission, showing the strong effect of this low temperature range on ortho and para H$_2$O rapidly changing from absorption to emission (or vice versa) in connecting transtions, according to Large Velocity Gradient models by \cite{2006ApJ...649L..33C}.   Narrow emission matching this prediciton is observed to just below 3$\sigma$, at a measured strength of 0.3 K km s$^{-1}$.  While just at a nominal confidence level observationally, the change from absorption to emission in the two transitions provides a good constraint on the gas temperature.

Since the 557 GHz profile is suggestive of CO with slightly enhanced red-shifted emission, the model there includes an extra component at $+75$ km s$^{-1}$ to roughly gauge the contribution where CO intensities peak in the NW of the Butterfly Nebula.  The asymmetry is largely unverified with such a contribution slightly above the noise level; excluding it leads to marginally higher (by 3-4\%) integrated intensities of the main emission and absorption lines.   

\begin{deluxetable*}{l r r r r r c r r}
  \tablewidth{0pc}
  \tablecaption{$\eta$ Car H$_2$O vapor properties}

 \tablehead{
 \colhead{Species and Transition} & 
\colhead{$\nu_0$} &
\colhead{$E_u$} &
\colhead{$\langle V_{\rm{LSR}} \rangle$} &
\colhead{$\Delta V$} & 
\colhead{$I_\nu$\tablenotemark{a}}  &
 \colhead{$I_\lambda$\tablenotemark{b}}  &
\colhead{$T_{\rm{gas}}$}  &
\colhead{$N$} \\
 \colhead{} & 
 \colhead{(GHz)} & 
\colhead{(K)} &
 \colhead{(km s$^{-1}$)} &
 \colhead{(km s$^{-1}$)} &
 \colhead{(K km s$^{-1}$)} &
 \colhead{(10$^{-17}$ W cm$^{-1}$)} &
\colhead{(K)} &
\colhead{(10$^{13}$ cm$^{-2}$)} 
 } 
 
 \startdata  

o-H$_2$O $1_{1,0} - 1_{0,1}$ & 556.936 & 61.0 & $+$8.6 & 160.0 & 4.0 (0.4) & 4.3 (1.0) & \dotfill & \dotfill \\
\multicolumn{3}{c}{\dotfill model comp. 1 - emission \dotfill} & $+$10.0 & 160.0 & 4.5 (0.8) &  6.0 (1.0) & 110 (25) & 4.80 (0.20) \\ 
\multicolumn{3}{c}{\dotfill model comp. 2 - absorption \dotfill } & $-$16.1 & 18.0 & 0.5 (0.1) &  2.0 (0.8) & 7.0 (2.0) & 1.30 (0.20) \\ 
p-H$_2$O $2_{0,2} - 1_{1,1}$ & 987.927 & 100.8 & $1$13.9 &165.0 &  5.8 (1.2) & 5.5 (2.0) & \dotfill & \dotfill \\
\multicolumn{3}{c}{\dotfill model comp. 1 - emission \dotfill} & $-$13.0 & 165.0 & 5.5 (0.8) &  5.0 (1.0) & 110 (25) & 2.10 (0.20) \\ 
\multicolumn{3}{c}{\dotfill model comp. 2 - emission \dotfill} & $-$15.0 & 15.0 & 0.3 (0.1) &  1.0 (0.5) & 110 (25) & 0.40 (0.15) \\ 

\tablenotetext{a}{The line intensity $I_\nu = \int T_{\rm{b}} dv$ integrated over the velocity range of the spectral feature on a main beam temperature scale, converted from antennae temperates using main beam sizes forward efficiencies of 38.$''$5 and 0.75 for the 1$_{1,0}$-1$_{0,1}$ line, and 21.$''$5 and 0.74 for the  2$_{0,2}$-1$_{1,1}$ line. The intensities are integrated over the composite observed feature, and over the individual broad and narrow components of the synthetic spectra.}
\tablenotetext{b}{A disk of diameter 1$''$ is assumed for the conversion from brightness temperature to irradiance units.}  

\enddata

\label{water}

\end{deluxetable*}

The prediction for the $p$-H$_2$O line also includes an interacting weak contribution by NH$_2$ at 987.2 GHz.  It is not presented as an identification or indication of NH$_2$, but is warranted to include here from the results of more complete study of the nitrogen hydrides being presented elsewhere \citep{2019MorrInPrep,2019GullInPrep}.  

In addition,  we show the model prediction of a potentially more relevant CH$_3$OH $5-6$ $\nu_t = 0\; E$ symmetry line ($E_u$ = 158.8 K) at 987.1024 GHz, based entirely on our simple model of the 231.3 GHz CH$_3$OH $10-9$ emission line.  The prediction of the 987 GHz line is $\sim$25\% too strong compared to the data, which is not unreasonable for the {\it{conservatively}} 2-$\sigma$ detection, more realistically 3-$\sigma$ (using $\sigma_{\rm{RMS}}$ = 0.025 K and $\Delta v_{\rm{FWHM}}$ = 75 km s$^{-1}$), but also limited by uncertainties in setting the surrounding baseline.  At the risk of over-interpretation, better agreement could be achieved if we were to treat the 987 GHz $A$ symmetry and 231 GHz $E$ symmetry lines as separate species, since they differ in their nuclear spin alignment and thus in energy scheme.  Reducing both excitation temperature and column density by 25-30\% leads to better fit of the 987 GHz line, and an $E:A$ abundance ratio of around 0.7,  which occurs at spin temperatures below 30 K \citep{2011A&A...533A..24W}.  We currently regard the feature at 987 GHz as useful upper limit to the strength on that emission line.  Further note that no strong absorption is detected here.  While an upper integrated intensity limit of $\sim$0.3 K km s$^{-1}$ could be placed on the excitation analysis of the absorbing gas, assuming it to be methanol, such a limit could be extremely misleading where our simple modeling is concerned in the low temperature regime in which non-LTE effects and the distinct energy band structure of the different $A$- and $E$-symmetry transitions must be considered in detail.  Even under ideal warm and dense molecular cloud conditions in SFRs, methonol poses and significant theoretical challenge, and is beyond the scope of this paper.   An excitation analysis of other methanol lines observable with ALMA will provide much better constraints on the identifications and physical conditions.

\subsection{The Water OPR}

The ortho-to-para water ratio (OPR) is sensitive to the gas temperature in thermal equilibrium due to the difference in rotational ground state energies of the two spin states.  Inelastic collisions do not affect the OPR, thus some clues about the water formation mechanism may be provided by this ratio.  The OPR reaches a warm temperature limit above 40 K to the spin statistical weights of 3; lower values occur at cooler temperatures, as formation of $p$-H$_2$O is favored over $o$-H$_2$O \citep{1987A&A...187..419M}.   

The OPR of the water emission in $\eta$ Car is estimated to be 2.3 $\pm$0.4, similar to that in Galactic interstellar clouds at corresponding spin temperatures of 27-32 K \citep{2013JPCA..117.9661L,2013ApJ...762...11F}.  This spin temperature is generally interpreted to correspond to some physical temperature at which the water molecules formed, either in the gas phase or on grain surfaces.  Hence the higher gas temperatures derived from our synthetic spectra are somewhat surprising (Table~\ref{water}).  However, laboratory experiments by \cite{2011JPCA..115.9682S} show that pure $p$-H$_2$O formed at 4 K rapidly switches spin state as the ice is gradually warmed to higher temperatures (to 260-280 K in those experiments), and the equilibrium OPR is reached in the vapor above the ice \citep{2013JPCA..117.9661L}. The spin conversion rate is already significant at 30 K.  In other words, the observed OPR retains no history of the formation temperature but instead reflects one (or a combination) of the current grain/ice temperature, or processing of water ice into vapor via thermal evaporation or photodesorption by UV photons. UV photodesorption would be favored at low column densities, as is clearly the case in $\eta$ Car.


\section{Discussion: Analogous Fischer-Tropsch-type synthesis of methanol in $\eta$ Car?} 

Here we address the question of how methanol may be formed in the environment of $\eta$ Car, assuming it to be the carrier of the 231 GHz feature.
Gas phase radiative association of CH$_3^+$ and H$_2$O could not lead to methanol formation here, since yields would be far too low, [CH$_3$OH/H$_2$] $\sim$ 10$^{-10}$ \citep{2006FaDi..133...51G}.  It is also unlikely to be present by desorption or sputtering from icy dust grain mantles --- the favored scenario for the ISM where methanol is a prominent ice component  \citep{2002ApJ...571L.173W,2009A&A...505..629F,2015ARA&A..53..541B} and may be non-thermally desorbed by UV radiation, cosmic rays, and/or various exothermic reactions, or sputtered by thermal heating as in the dense molecular envelopes of YSOs \citep{2009ARA&A..47..427H,2010A&A...516A..57K}.  From an observational standpoint, mid-IR observations where we would search for  CO, CO$_2$, H$_2$O, etc., ices do not yet resolve the central region spectrally and spatially at the necessary wavelengths.  However, the detection of H$_2$O in $\eta$ Car  (Sec.~\ref{watersection}) means that ice mantle removal cannot be the source of the detected CH$_3$OH.  The measured  CH$_3$OH/H$_2$O abundance ratio is greater than 10$^3$, whereas desorption of interstellar ice mantles would produce ratios typically of a few percent \citep{2015ARA&A..53..541B}. Furthermore, the radiation environment of $\eta$ Car  leads to short lifetimes for desorbed CH$_3$OH molecules due to photodissociation and ionization.  Methanol will be protonated by HCO$^+$ and N$_2$H$^+$  and electron dissociative recombination of CH$_3$OH$_2^+$  results in  fragmentation in $> 95\%$ of  reactions \citep{2006FaDi..133..177G}.  Thus, the observed methanol cannot have its origin in a one-time ice desorption event and instead must be continuously replenished. 

In $\eta$ Car, the presence of CO and H$_2$, and at elevated gas and dust temperatures ($ \sim 100-800$ K) and densities, a promising mechanism for CH$_3$OH production is Fischer-Tropsch-type (FTT) catalysis on dust \citep{1991Christ}. Although the pressures in $\eta$ Car are far below those encountered in laboratory experiments set for commercial applications, FTT processes can be viable since the relative  timescales available are on the order of  centuries to several millennia, as compared to days \citep{2008ApJ...673L.225N}.  In FTT catalysis CO and H$_2$ molecules can adsorb either dissociatively or non-dissociatively on a hot substrate and initiate  a sequence surface reactions whose products depend on the physical conditions (pressure, temperature) and the nature of the substrate.  

There are three overall reactions of interest.  The first is {\it `methanation' }
              
\begin{equation}
{\rm CO ~+~ 3H_2 ~\rightleftharpoons~  CH_4   ~+~ H_2O}\label{eq:ch4}
\end{equation}

\noindent which can also produce higher hydrocarbons depending on the  reaction conditions. The second is {\it methanol synthesis }

\begin{equation}
{\rm CO ~+~ 2H_2 ~\rightleftharpoons~  CH_3OH  }\label{eq:meth}
\end{equation}

\noindent and the third is {\it oxidation }
 
\begin{equation}
{\rm 4CO ~+~ 2H_2 ~\rightleftharpoons~  CO_2 ~+~ 2H_2O ~+~ 3C  }\label{eq:coke}
\end{equation}

\noindent In each process the product molecules are desorbed into the gas phase, except in reaction (\ref{eq:coke}) where the C atoms can react to build  
 carbonaceous layers on the substrate,  `poisoning'  surface  binding sites and  leading to a reduction of the catalytic efficiency.   

The solid materials identified  as  plausible components of the circumstellar dust  surrounding $\eta$ Car \citep{2017ApJ...842...79M} could potentially act as quasi-FTT catalysts.  Among these catalysts in industrial applications, metallic Fe substrates are the best-known. Fe dust grains would tend to favor formation of CH$_4$ and higher hydrocarbons along with water through reaction (\ref{eq:ch4}), with only small amounts of methanol produced.  The low H$_2$O abundance detected  (to an upper limit of $\sim  10^{-9}$; see below) is far below that expected and could be explained if in fact the Fe dust has already built up a carbonaceous covering through reaction (\ref{eq:coke}), such that catalysis via reaction (\ref{eq:ch4}) is significantly reduced.
 
On the other hand, reaction  (\ref{eq:meth})  could in principle produce large abundances of CH$_3$OH and no water, depending on whether suitable substrates are available; industrially these are typically oxides of Zn, Cu and Cr. Most of  the putatively-identified dust materials in $\eta$ Car (e.g., Fe, Al$_2$O$_3$, nitrides, pyroxene and metal-containing silicates) have not been evaluated for their efficacy in  FTT synthesis as they are unlikely to be efficient catalysts on the commercial scale.  Although FTT catalysis by Fe-containing silicate particles as proxies for solar nebula dust has been studied  \citep{2008ApJ...673L.225N},  gaseous CH$_3$OH was not identified as a reaction product in subsequent analyses \citep{2016M&PS...51.1310N}.  It therefore remains to be determined by experiment  which of  the solid particles produced in the peculiar chemistry of $\eta$ Car can exhibit the  necessary selectivity for methanol synthesis.

\section{Summary and Conclusions}

ALMA observations of the central region in $\eta$ Car at 231-232 GHz, taken near binary orbit phase $\Delta \phi$ = 0.84, depict an intensity distribution of CO $J=2-1$ tracing molecular layers of  the disrupted, massive equatorial torus, generally consistent with observations taken in the same epoch over a few months following periastron \citep{2018MNRAS.474.4988S}, and with CO $J=4-3$ observations taken mid-orbit \citep{2019MNRAS.490.1570B}.  Minor differences in morphology between these sets of observations can be partly attributed to differences in radiometric and angular sensitivity, but also some potential averaging-out of local intensity variations in the earlier post-periastron observations by a mix of antenna configurations.   These results provide modeling constraints on temperature, ionization, shielding, and excitation response times with additional ALMA observations planned during periastron passage, when the emerging UV field from the hot secondary star will be suppressed in the stellar wind of the primary.      

In contrast to the CO brightness distribution, the carrier of a previously unidentified feature at 231.3 GHz is confined to the central region, less than 2$''$ (0.02 pc) across, and exhibits emission and prominent absorption components will little profile variation over the observed region.  We attribute both components to CH$_3$OH 10$_2 - 9_3 \; A^-$, well-centered at 231.2811 GHz and formed in proximity to the central source.  We prefer this identification over $^{13}$CS $J=5-4$ (which would also be a first detection), blue-shifted by $-$80 km s$^{-1}$, on distribution and velocity structure arguments in relation to compact fast-moving material traced by H emission, HCN and H$^{13}$CN absorption spectra \citep{2014ApJ...791...95A,2020AbrahamInPrep,2019MNRAS.490.1570B}.   Other transitions of methanol and CS occur at frequencies accessible to ALMA for future confirmation. 

Motivated by the (probable) methanol discovery, we have further identified two water transitions in earlier unpublished {\it{Herschel}}/HIFI observations, in ortho and para spin states at 557 GHz and 988 GHz.  Both water and methanol are first detections around a high-mass star system in its final phases evolution, verging on a possible Type II or Ib/c supernova.  HIFI does not provide any spatial information due to the large beam, but the velocity structure of the two lines suggest they are formed close to the central source as well. 

The cool methanol is optically thin to the sub-mm continuum, but may contribute a considerable amount of extinction of UV and optical radiation from the star in the line of sight.  Significant photometric variability in the thermal IR is clearly evident in mainly large-aperture observations compiled over 50 years, but not discernibly on a long-term trend \citep{2019A&A...630L...6M}.  Periodicity of thermal emission from the central knot versus the surrounding nebula, and in relation to orbit phase, is difficult to establish due to inadequate temporal sampling and the generally large apertures.  Furthermore, while we are convinced of an association between the warm and cool methanol to the central region rather than formation in the ISM, the relationship to other compact structures in the line of sight is unknown.  We expect to gain additional insights on these relationships through monitoring observations with ALMA, particularly during periastron passage when we can gage light travel and response times compared to other locations in the nebula.  

A focus on water in this environment is also highly desirable to strengthen the detections in the HIFI spectra, and further constrain parameters describing the physical conditions in the water vapor cloud.  Unfortunately follow-up of low excitation water will not be possible until instruments on the next generation of space telescopes such as the Origins Space Telescope or the Space Infrared Telescope for Cosmology and Astrophysics (SPICA) are in operation.  Nonetheless, the range of excitation temperatures and thus column densities are reasonable (to withing the quoted uncertainties) due to temperature sensitivity of the gas leading to emission or absorption.   

The very high abundance ratio [CH$_3$OH/H$_2$O] largely precludes desorption or sputtering of icy grain mantles as the origin of methanol gas. A more promising pathway is through processes analogous to FTT, possibly catalyzed by metals or metal-bearing dust grains to keep water production to observationally low levels. More laboratory work is needed to identify components of $\eta$ Car's dust that may be more suitable catalysts for methanol synthesis, and at much lower pressures than demanded by FTT processes for efficient production of commercial petroleum products.

Although $\eta$ Car is one of the most unique star systems in the Galaxy, the conditions fostering production and survival of complex hydrocarbons and water may not be unique to its circumstellar environment.  Within the last decade, it has become observationally evident that O-type stars with sufficient initial initial mass, $\gtrsim$ 20 $M_\odot$, to evolve into progenitors of core collapse supernova are born in close binaries or multiple systems, and thus experience interactions between companions that can dominate their evolution \citep{2012Sci...337..444S,2017IAUS..329..110S,2019arXiv190706687Z}.  As a consequence, a sizable fraction of massive stars experience colliding wind and mass transfer effects, and may ultimately result in mergers under a number of conditions following orbit properties, stellar mass ratios, and rotation rates \citep{2013EAS....64...21D,2014ApJ...782....7D,2017ApJS..230...15M}.   Merger events in a contact phase between companions as one leaves the core-H burning main sequence \citep{2014ApJ...796..121J}, and by collisions in dense stellar clusters \citep{2013MNRAS.434.3497G} attempt to explain the main observational properties of LBVs as a class.   

While there remain large gaps in our knowledge of $\eta$ Car as a likely merger from an initially hierarchical triple system \citep{2016MNRAS.456.3401P,2018MNRAS.480.1466S}, the presence of light molecules with wide velocity dispersions may be a signpost of a massive merger remnant, where the observed distribution or derived physical conditions of the molecular gas rule out association with cooling shocked gas from wind-wind collisions (i.e., not arising from a fast wind having overtaken a previous slower wind) or from a wind-ISM interface.  Complex molecules formed closer to the system, by contrast, could trace a more compact, dense colliding wind structure, prior to depletion of H needed for hydrogenization of CO in the case of methanol. Single stars in the carbon-rich WC class of Wolf-Rayet stars can form carbonaceous dust in densely clumped winds, for example \citep{2012MNRAS.426.1720D}, but are unable to form hydrocarbons due to the H depletion that defines their spectral class and evolutionary state.  Binary stars with an OB-type companion, however, provide a more chemically suitable environment.   A number of nearby LBV stars and Wolf-Rayet binaries may be ideal for testing these ideas.  The abundances would be predictably low, presenting a stimulating and potentially rewarding challenge for the next generation of far-IR and sub-mm instruments.  

\begin{center}
{\sc{Acknowledgments}}
\end{center}

The authors express their appreciation to Pierre Cox, Zulema Abraham, and Pedro Paulo Bonetti Beaklini for productive discussions on the ALMA observations, and Darius Liz for cautionary words on the water OPR.  We also thank our anonymous referee for comments leading to important improvements in the manuscript.  This paper makes use of the following ALMA data: ADS/JAO.ALMA \#2018.A.00026.S and ADS/JAO.ALMA \#2013.1.00661.S. ALMA is a partnership of ESO (representing its member states), NSF (USA) and NINS (Japan), together with NRC (Canada), MOST and ASIAA (Taiwan), and KASI (Republic of Korea), in cooperation with the Republic of Chile. The Joint ALMA Observatory is operated by ESO, AUI/NRAO and NAOJ.  The National Radio Astronomy Observatory is a facility of the National Science Foundation operated under cooperative agreement by Associated Universities, Inc.  We have also made use of observations obtained from the {\it{Herschel}} Space Observatory.  {\it{Herschel}} is an ESA space observatory with science instruments provided by European-led Principal Investigator consortia and with important participation from NASA.  AD thanks to the National Concil for Scientific and Technological Development (CNPq) and to Funda\c{c}\~ao de  Amparo\'a  Pesquisa  do  Estado  de  S\~ao  Paulo  (FAPESP)  for continuing support.


\begin{thebibliography}{}

\bibitem[Abraham et al.(2014)]{2014ApJ...791...95A} Abraham, Z., Falceta-Gon{\c c}alves, D., \& Beaklini, P.~P.~B.\ 2014, \apj, 791, 95 
\bibitem[Abraham et al.(2020)]{2020AbrahamInPrep} Abraham, Z., et al., in preparation.
\bibitem[Bergin \& Snell(2002)]{2002ApJ...581L.105B} Bergin, E.~A., \& Snell, R.~L.\ 2002, \apjl, 581, L105
\bibitem[Bolatto et al.(2013)]{2013ARA&A..51..207B} Bolatto, A.~D., Wolfire, M., \& Leroy, A.~K.\ 2013, \araa, 51, 207
\bibitem[Boogert et al.(2015)]{2015ARA&A..53..541B} Boogert, A.~C.~A., Gerakines, P.~A., \& Whittet, D.~C.~B.\ 2015, Annual Review of Astronomy and Astrophysics, 53, 541
\bibitem[Boonman et al.(2003)]{2003A&A...406..937B} Boonman, A.~M.~S., Doty, S.~D., van Dishoeck, E.~F., et al.\ 2003, \aap, 406, 937
\bibitem[Bordiu \& Rizzo(2019)]{2019MNRAS.490.1570B} Bordiu, C., \& Rizzo, J.~R.\ 2019, \mnras, 490, 1570
\bibitem[Breen et al.(2013)]{2013MNRAS.435..524B} Breen, S.~L., Ellingsen, S.~P., Contreras, Y., et al.\ 2013, \mnras, 435, 524
\bibitem[Brooks et al.(1998)]{1998PASA...15..202B} Brooks, K.~J., Whiteoak, J.~B., \& Storey, J.~W.~V.\ 1998, \pasa, 15, 202
\bibitem[Cernicharo et al.(2006)]{2006ApJ...649L..33C} Cernicharo, J., Goicoechea, J.~R., Daniel, F., et al.\ 2006, \apjl, 649, L33
\bibitem[Chesneau et al.(2005)]{2005ASPC..332..165C} Chesneau, O., Min, M., Herbst, T., et al.\ 2005, The Fate of the Most Massive Stars, 167
\bibitem[Christman(1991)]{1991Christ} Christmann, K. 1991. {\it Introduction to Surface Physical Chemistry}, Springer-Verlag, New York, New York, USA
\bibitem[Corcoran et al.(2017)]{2017ApJ...838...45C} Corcoran, M.~F., Liburd, J., Morris, D., et al.\ 2017, \apj, 838, 45
\bibitem[Cox \& Bronfman(1995)]{1995A&A...299..583C} Cox, P., \& Bronfman, L.\ 1995, \aap, 299, 583
\bibitem[Damineli et al.(2008a)]{2008MNRAS.384.1649D} Damineli, A., Hillier, D.~J., Corcoran, M.~F., et al.\ 2008a, \mnras, 384, 1649 
\bibitem[Damineli et al.(2008b)]{2008MNRAS.386.2330D} Damineli, A., Hillier, D.~J., Corcoran, M.~F., et al.\ 2008b \mnras, 386, 2330
\bibitem[Damineli et al.(2019)]{2019MNRAS.484.1325D} Damineli, A., Fern{\'a}ndez-Laj{\'u}s, E., Almeida, L.~A., et al.\ 2019, \mnras, 484, 1325 
\bibitem[David-Uraz et al.(2012)]{2012MNRAS.426.1720D} David-Uraz, A., Moffat, A.~F.~J., Chen{\'e}, A.-N., et al.\ 2012, \mnras, 426, 1720
\bibitem[Davidson \& Humphreys(1997)]{1997ARA&A..35....1D} Davidson, K., \& Humphreys, R.~M.\ 1997, \araa, 35, 1 
\bibitem[Davidson et al.(2017)]{2017RNAAS...1a...6D} Davidson, K., Ishibashi, K., \& Martin, J.~C.\ 2017, Research Notes of the American Astronomical Society, 1, 6
\bibitem[Dickel \& Wall(1974)]{1974A&A....31....5D} Dickel, H.~R., \& Wall, J.~V.\ 1974, \aap, 31, 5
\bibitem[Emprechtinger et al.(2013)]{2013ApJ...765...61E} Emprechtinger, M., Lis, D.~C., Rolffs, R., et al.\ 2013, \apj, 765, 61
\bibitem[Flagey et al.(2013)]{2013ApJ...762...11F} Flagey, N., Goldsmith, P.~F., Lis, D.~C., et al.\ 2013, \apj, 762, 11
\bibitem[Fuchs et al.(2009)]{2009A&A...505..629F} Fuchs, G.~W., Cuppen, H.~M., Ioppolo, S., et al.\ 2009, \aap, 505, 629
\bibitem[Gardner et al.(1973)]{1973A&A....23...51G} Gardner, F.~F., Dickel, H.~R., \& Whiteoak, J.~B.\ 1973, \aap, 23, 51
\bibitem[Garrod et al.(2006)]{2006FaDi..133...51G} Garrod, R., Park, I.~H., Caselli, P., et al.\ 2006, Faraday Discussions, 133, 51
\bibitem[Geppert et al.(2006)]{2006FaDi..133..177G} Geppert, W.~D., Hamberg, M., Thomas, R.~D., et al.\ 2006, Faraday Discussions, 133, 177
\bibitem[Glebbeek et al.(2013)]{2013MNRAS.434.3497G} Glebbeek, E., Gaburov, E., Portegies Zwart, S., et al.\ 2013, \mnras, 434, 3497
\bibitem[de Graauw et al.(1996)]{1996A&A...315L..49D} de Graauw, T., Haser, L.~N., Beintema, D.~A., et al.\ 1996, \aap, 315, L49
\bibitem[Griffin et al.(2010)]{2010A&A...518L...3G} Griffin, M.~J., Abergel, A., Abreu, A., et al.\ 2010, \aap, 518, L3
\bibitem[Groh et al.(2012)]{2012MNRAS.423.1623G} Groh, J.~H., Hillier, D.~J., Madura, T.~I., et al.\ 2012, \mnras, 423, 1623
\bibitem[Gull et al.(2016)]{2016MNRAS.462.3196G} Gull, T.~R., Madura, T.~I., Teodoro, M., et al.\ 2016, \mnras, 462, 3196 
\bibitem[Gull et al.(2020)]{2019GullInPrep} Gull, T.~R., et al., in preparation. 
\bibitem[Hamaguchi et al.(2018)]{2018NatAs...2..731H} Hamaguchi, K., Corcoran, M.~F., Pittard, J.~M., et al.\ 2018, Nature Astronomy, 2, 731
\bibitem[Herbst \& van Dishoeck(2009)]{2009ARA&A..47..427H} Herbst, E., \& van Dishoeck, E.~F.\ 2009, Annual Review of Astronomy and Astrophysics, 47, 427
\bibitem[Herpin et al.(2012)]{2012A&A...542A..76H} Herpin, F., Chavarr{\'\i}a, L., van der Tak, F., et al.\ 2012, \aap, 542, A76
\bibitem[Justham et al.(2014)]{2014ApJ...796..121J} Justham, S., Podsiadlowski, P., \& Vink, J.~S.\ 2014, \apj, 796, 121
\bibitem[Kristensen et al.(2010)]{2010A&A...516A..57K} Kristensen, L.~E., van Dishoeck, E.~F., van Kempen, T.~A., et al.\ 2010, \aap, 516, A57
\bibitem[de Koter et al.(2013)]{2013EAS....64...21D} de Koter, A., Bestenlehner, J.~M., de Mink, S.~E., et al.\ 2013, EAS Publications Series, 21
\bibitem[Lis et al.(2013)]{2013JPCA..117.9661L} Lis, D.~C., Bergin, E.~A., Schilke, P., et al.\ 2013, Journal of Physical Chemistry A, 117, 9661
\bibitem[Loinard et al.(2012)]{2012ApJ...749L...4L} Loinard, L., Menten, K.~M., G{\"u}sten, R., Zapata, L.~A., \& Rodr{\'{\i}}guez, L.~F.\ 2012, \apjl, 749, L4 
\bibitem[Loinard et al.(2016)]{2016ApJ...833...48L} Loinard, L., Kami{\'n}ski, T., Serra, P., et al.\ 2016, \apj, 833, 48
\bibitem[Madura \& Groh(2012)]{2012ApJ...746L..18M} Madura, T.~I., \& Groh, J.~H.\ 2012, \apjl, 746, L18
\bibitem[Madura et al.(2012)]{2012MNRAS.420.2064M} Madura, T.~I., Gull, T.~R., Owocki, S.~P., et al.\ 2012, \mnras, 420, 2064
\bibitem[Madura et al.(2013)]{2013MNRAS.436.3820M} Madura, T.~I., Gull, T.~R., Okazaki, A.~T., et al.\ 2013, \mnras, 436, 3820
\bibitem[Mehner et al.(2010)]{2010ApJ...710..729M} Mehner, A., Davidson, K., Ferland, G.~J., et al.\ 2010, \apj, 710, 729
\bibitem[Mehner et al.(2019)]{2019A&A...630L...6M} Mehner, A., de Wit, W.-J., Asmus, D., et al.\ 2019, \aap, 630, L6
\bibitem[de Mink et al.(2014)]{2014ApJ...782....7D} de Mink, S.~E., Sana, H., Langer, N., et al.\ 2014, \apj, 782, 7
\bibitem[Moe \& Di Stefano(2017)]{2017ApJS..230...15M} Moe, M., \& Di Stefano, R.\ 2017, \apjs, 230, 15
\bibitem[Moffat \& Corcoran(2009)]{2009ApJ...707..693M} Moffat, A.~F.~J., \& Corcoran, M.~F.\ 2009, \apj, 707, 693
\bibitem[Morris et al.(1999)]{1999Natur.402..502M} Morris, P.~W., Waters, L.~B.~F.~M., Barlow, M.~J., et al.\ 1999, \nat, 402, 502
\bibitem[Morris(2015)]{2015wrs..conf..155M} Morris, P.~W.\ 2015, Wolf-Rayet Stars: Proceedings of an International Workshop Held in Potsdam, 155
\bibitem[Morris et al.(2020)]{2019MorrInPrep} Morris, P.~W., et al., in preparation.
\bibitem[Morris et al.(2017)]{2017ApJ...842...79M} Morris, P.~W., Gull, T.~R., Hillier, D.~J., et al.\ 2017, \apj, 842, 79
\bibitem[Morse et al.(2001)]{2001ApJ...548L.207M} Morse, J.~A., Kellogg, J.~R., Bally, J., et al.\ 2001, \apjl, 548, L207 
\bibitem[Mumma et al.(1987)]{1987A&A...187..419M} Mumma, M.~J., Weaver, H.~A., \& Larson, H.~P.\ 1987, \aap, 187, 419
\bibitem[Neill et al.(2014)]{2014ApJ...789....8N} Neill, J.~L., Bergin, E.~A., Lis, D.~C., et al.\ 2014, \apj, 789, 8
\bibitem[Nuth et al.(2008)]{2008ApJ...673L.225N} Nuth, J.~A., Johnson, N.~M., \& Manning, S.\ 2008, \apj, 673, L225
\bibitem[Nuth et al.(2016)]{2016M&PS...51.1310N} Nuth, J.~A., Johnson, N.~M., Ferguson, F.~T., et al.\ 2016, Meteoritics and Planetary Science, 51, 1310
\bibitem[Olofsson et al.(2017)]{2017A&A...603L...2O} Olofsson, H., Vlemmings, W.~H.~T., Bergman, P., et al.\ 2017, \aap, 603, L2
\bibitem[Olofsson et al.(2019)]{2019A&A...623A.153O} Olofsson, H., Khouri, T., Maercker, M., et al.\ 2019, \aap, 623, A153
\bibitem[Parfenov et al.(2016)]{2016MNRAS.460.2648P} Parfenov, S.~Y., Semenov, D.~A., Sobolev, A.~M., et al.\ 2016, \mnras, 460, 2648
\bibitem[Pickett et al.(1998)]{1998JQSRT..60..883P} Pickett, H.~M., Poynter, R.~L., Cohen, E.~A., et al.\ 1998, \jqsrt, 60, 883; https://spec.jpl.nasa.gov/
\bibitem[Pittard \& Corcoran(2002)]{2002A&A...383..636P} Pittard, J.~M., \& Corcoran, M.~F.\ 2002, \aap, 383, 636
\bibitem[Polomski et al.(1999)]{1999AJ....118.2369P} Polomski, E.~F., Telesco, C.~M., Pi{\~n}a, R.~K., \& Fisher, R.~S. 1999, \aj, 118, 2369
\bibitem[Portegies Zwart \& van den Heuvel(2016)]{2016MNRAS.456.3401P} Portegies Zwart, S.~F., \& van den Heuvel, E.~P.~J.\ 2016, \mnras, 456, 3401
\bibitem[Pottage et al.(2004)]{2004MNRAS.352...39P} Pottage, J.~T., Flower, D.~R., \& Davis, S.~L.\ 2004, \mnras, 352, 39
\bibitem[Quintana-Lacaci et al.(2016)]{2016A&A...592A..51Q} Quintana-Lacaci, G., Ag{\'u}ndez, M., Cernicharo, J., et al.\ 2016, \aap, 592, A51
\bibitem[Rabli \& Flower(2010)]{2010MNRAS.403.2033R} Rabli, D., \& Flower, D.~R.\ 2010, \mnras, 403, 2033
\bibitem[Remmen et al.(2013)]{2013ApJ...773...27R} Remmen, G.~N., Davidson, K., \& Mehner, A.\ 2013, \apj, 773, 27
\bibitem[Roelfsema et al.(2012)]{2012A&A...537A..17R} Roelfsema, P.~R., Helmich, F.~P., Teyssier, D., et al.\ 2012, \aap, 537, A17
\bibitem[Salii \& Sobolev(2006)]{2006ARep...50..965S} Salii, S.~V., \& Sobolev, A.~M.\ 2006, Astronomy Reports, 50, 965
\bibitem[Sana(2017)]{2017IAUS..329..110S} Sana, H.\ 2017, The Lives and Death-throes of Massive Stars, 110
\bibitem[Sana et al.(2012)]{2012Sci...337..444S} Sana, H., de Mink, S.~E., de Koter, A., et al.\ 2012, Science, 337, 444
\bibitem[S{\'a}nchez Contreras et al.(2018)]{2018A&A...618A.164S} S{\'a}nchez Contreras, C., Alcolea, J., Bujarrabal, V., et al.\ 2018, \aap, 618, A164
\bibitem[Sliter et al.(2011)]{2011JPCA..115.9682S} Sliter, R., Gish, M., \& Vilesov, A.~F.\ 2011, Journal of Physical Chemistry A, 115, 9682
\bibitem[Smith(2004)]{2004MNRAS.351L..15S} Smith, N.\ 2004, \mnras, 351, L15 
\bibitem[Smith(2006)]{2006ApJ...644.1151S} Smith, N.\ 2006, \apj, 644, 1151
\bibitem[Smith \& Frew(2011)]{2011MNRAS.415.2009S} Smith, N., \& Frew, D.~J.\ 2011, \mnras, 415, 2009 
\bibitem[Smith et al.(2018a)]{2018MNRAS.474.4988S} Smith, N., Ginsburg, A., \& Bally, J.\ 2018, \mnras, 474, 4988
\bibitem[Smith et al.(2018b)]{2018MNRAS.480.1466S} Smith, N., Andrews, J.~E., Rest, A., et al.\ 2018, \mnras, 480, 1466
\bibitem[Snell et al.(2000)]{2000ApJ...539L.101S} Snell, R.~L., Howe, J.~E., Ashby, M.~L.~N., et al.\ 2000, \apjl, 539, L101
\bibitem[Steffen et al.(2014)]{2014MNRAS.442.3316S} Steffen, W., Teodoro, M., Madura, T.~I., et al.\ 2014, Monthly Notices of the Royal Astronomical Society, 442, 3316
\bibitem[van der Tak et al.(2007)]{2007A&A...468..627V} van der Tak, F.~F.~S., Black, J.~H., Sch{\"o}ier, F.~L., Jansen, D.~J., \& van Dishoeck, E.~F.\ 2007, \aap, 468, 627 
\bibitem[Teodoro et al.(2016)]{2016ApJ...819..131T} Teodoro, M., Damineli, A., Heathcote, B., et al.\ 2016, \apj, 819, 131 
\bibitem[Watanabe \& Kouchi(2002)]{2002ApJ...571L.173W} Watanabe, N., \& Kouchi, A.\ 2002, \apjl, 571, L173
\bibitem[Weigelt \& Ebersberger(1986)]{1986A&A...163L...5W} Weigelt, G., \& Ebersberger, J.\ 1986, \aap, 163, L5
\bibitem[Wirstr{\"o}m et al.(2011)]{2011A&A...533A..24W} Wirstr{\"o}m, E.~S., Geppert, W.~D., Hjalmarson, {\r{A}}., et al.\ 2011, \aap, 533, A24
\bibitem[Zapartas et al.(2019)]{2019arXiv190706687Z} Zapartas, E., de Mink, S.~E., Justham, S., et al.\ 2019, arXiv e-prints, arXiv:1907.06687
\bibitem[Zethson et al.(2012)]{2012A&A...540A.133Z} Zethson, T., Johansson, S., Hartman, H., et al.\ 2012, \aap, 540, A133


\end{thebibliography}
\end{document}